\documentclass[aps,pra,twocolumn,groupedaddress,amsfonts,showpacs]{revtex4}
\usepackage{epsfig}
\usepackage{bm}

\begin{document}

\title{Classical light dispersion theory in a regular lattice}

\author{M. Marino}
\email[]{Massimo.Marino@unimi.it}
\author{A. Carati}
\email[]{Andrea.Carati@unimi.it}
\author{L. Galgani}
\email[]{Luigi.Galgani@unimi.it}

\affiliation{Dipartimento di Matematica, Universit\`{a} di Milano,
via Saldini 50, I-20133 Milano (Italy)}

\date{November 10, 2006}

\begin{abstract}
We study the dynamics of an infinite regular lattice of classical
charged oscillators. Each individual oscillator is described as a
point particle subject to a harmonic restoring potential, to the
retarded electromagnetic field generated by all the other
particles, and to the radiation reaction expressed according to
the Lorentz--Dirac equation. Exact normal mode solutions,
describing the propagation of plane electromagnetic waves through
the lattice, are obtained for the complete linearized system of
infinitely many oscillators. At variance with all the available
results, our method is valid for any values of the frequency, or
of the ratio between wavelength and lattice parameter. A
remarkable feature is that the proper inclusion of radiation
reaction in the dynamics of the individual oscillators does not
give rise to any extinction coefficient for the global normal
modes of the lattice. The dispersion relations resulting from our
solution are numerically studied for the case of a simple cubic
lattice. New predictions are obtained in this way about the
behavior of the crystal at frequencies near the proper oscillation
frequency of the dipoles.
\end{abstract}
 \maketitle

\section{Introduction}

The classical theory of dispersion is a subject with a long and
noble history \cite{born1,BW}. Although the main features of the
phenomenon can be described by treating matter as a continuum
characterized by macroscopic quantities such as the electric and
magnetic polarizations, it is clear that a truly fundamental
theory has to be based on a microscopic model of matter. We shall
now try to summarize some crucial aspects of the problem in an
historical perspective, before illustrating the new features of
our present approach.

By treating an elementary electric dipole as an oscillator subject
to a linear restoring force, it is possible to obtain a simple
expression for the molecular polarizability, i.e.\ the complex
frequency-dependent linear coefficient which relates the
microscopic dipole moment to the amplitude of the incident
electromagnetic radiation. In order to correctly apply this simple
model to the description of the behavior of a large system of
mutually interacting dipoles, one has however to consider that the
field acting on each microscopic oscillator cannot be simply
identified with the macroscopic electromagnetic field in the
medium. In fact, while the latter simply represents the average of
the microscopic field over a region much larger than the
intermolecular spacing, the former has to be carefully calculated
by evaluating and summing, on the site occupied by the considered
dipole, the retarded fields generated by all the other dipoles of
the medium. This ``exciting'' field (as we shall refer to in the
following, although the names ``effective'' or ``local'' field
have also been employed in the literature) was theoretically
estimated by Lorentz already at the end of nineteenth century
\cite{lorentz1,lorentz2,lorentz3,lorentz4} by dividing the medium
into two regions separated by a virtual sphere surrounding the
considered dipole. He restricted his attention to the typical
situation in which the wavelength of the macroscopic
electromagnetic field is of a larger order of magnitude than the
average intermolecular spacing, so that one can take for the
virtual sphere a radius intermediate between the two. He then
argued that the influence of the portion of the medium lying
outside the sphere can be fairly approximated as that of a
continuous distribution of electric dipole moment, whereas the sum
of the forces exerted by all the dipoles situated inside the
sphere can be assumed to vanish in most cases on the basis of
symmetry considerations. A rather similar analysis, leading to
equivalent conclusions, was also carried out by Planck
\cite{planck}. With these arguments one can derive the well-known
Lorentz--Lorenz formula \cite{lorentz2,lorenz}, relating the
macroscopic dielectric constant of the medium to the molecular
polarizability, and it is thus possible to deduce an approximate
expression for the dispersion relation of an array of oscillators
in the long-wavelength regime (which generally includes the
optical frequencies).

A detailed microscopic theory of dispersion in a crystalline
solid, although with neglect of radiation reaction, was formulated
by Ewald \cite{ewald2,ewald1}. He considered a rectangular
parallelepiped as the unit cell of the Bravais lattice, and his
results were subsequently generalized by Born to more general
crystal structures \cite{born2,born1}. The mathematical methods
used by these authors (one has to keep in mind that the theory of
distributions was not yet existing at that time) led however to
rather clumsy expressions for the exciting field, which could be
numerically evaluated only in the limit of an infinitely large
ratio between wavelength and lattice constant, i.e.\ still
essentially in the continuum approximation. In this way the
previous results by Lorentz and Planck were recovered for
structures with tetrahedral symmetry. Furthermore, in the case of
parallelepipeds of unequal edges, Ewald was able to perform in the
same limit a numerical calculation relating the ratio between the
edges to the phenomenon of double refraction. Finally, Ewald
extended his analysis of the model also to the study of X-ray
diffraction \cite{ewald3}, but he made use to this purpose of
other important simplifications which are possible only in the
opposite limit of a radiation frequency much higher that the
characteristic frequencies of the crystal.

Many investigations were later devoted to the application of
quantum mechanics to the theory of light dispersion, and the
results of Ewald and Born were apparently considered to be the
final word about the problem of the mutual interaction of a large
array of classical resonators. We are going to prove that, on the
contrary, a deeper analysis reveals important properties of this
fundamental dynamical system which have been for many decades
completely overlooked.

In the present paper we shall study a system of infinitely many
charged particles, subject to linear restoring forces towards
their equilibrium positions at the sites of a regular lattice, and
interacting with each other through the retarded electromagnetic
fields. Our outset will therefore be similar to Ewalds's, but with
inclusion in the equations of motion of the usual ``triple-dot''
radiation reaction term, which corresponds to the nonrelativistic
form of the Lorentz--Dirac equation \cite{lorentz4,dirac}. The
only approximation that we shall use is that of small
oscillations: this will allow us still to deal with a system of
linear equations. We shall provide a general and rigorous
procedure for the calculation of the exciting field, avoiding to
introduce at any stage of the procedure the continuum
approximation. This will be accomplished by a method involving the
careful subtraction of two divergent quantities (representing
respectively the total field and the field generated by the dipole
under consideration), which appears to be more powerful than that
used by Ewald and Born, and presents some formal analogy with the
renormalization techniques of quantum field theory. Using our
procedure we shall show that for an infinite regular lattice there
exists a continuous set of normal modes which describe the
propagation of plane electromagnetic waves.

A remarkable result will be that the inclusion of the radiation
damping term in the equations of motion of the oscillating
particles, instead of giving rise to an extinction coefficient for
the wave, as is commonly believed according to the standard
approximated treatments of the model, is on the contrary essential
for justifying the presence of undamped collective waves. Such a
result in fact constitutes an extension to the three dimensional
case of an analogous one already obtained by two of the present
authors for the case of a rectilinear chain of one-dimensional
oscillators \cite{CG}. It relies upon a remarkable identity which
was originally formulated in a different context by Wheeler and
Feynman \cite{WF}. These authors deduced it from the hypothesis of
the ``complete absorber'', which they introduced in order make
their time-reversible action-at-a-distance electrodynamics
compatible with the observed phenomenon of radiation reaction. For
the physical system here considered we are going to prove in a
simple and direct way that, although no absorption mechanism is
present in the model, this ``Wheeler--Feynman identity'' actually
holds as a purely mathematical property of the entire class of
solutions on which we are interested.

By numerically studying the dispersion relations for the crystal,
as resulting from the exact solutions of the model, it will also
be shown that completely new features appear for frequencies in a
region about the proper frequency of the oscillators. In such a
region the wavelength can in fact become as short as the lattice
spacing, so that the approximations adopted in the previous
literature become unavailable, and only an exact solution can give
predictions about the behavior of the system. It is found that
inside the interval of frequencies where undamped wave propagation
was believed to be impossible, plane waves can actually propagate,
with very low group velocities, along certain lattice directions
and for appropriate wave polarization.

\section{The model}

Let us consider an infinite three-dimensional simple Bravais lattice,
that is an array of points
\begin{equation}
{\bf r}_{{\bf n}}=n_{1}{\bf a}_{1}+n_{2}{\bf a}_{2}+n_{3}{\bf a}_{3}\,,
\label{lattice}
\end{equation}
where ${\bf n}$ denotes the triple of relative
integers $(n_{1},n_{2},n_{3})$,
and ${\bf a}_{1}$, ${\bf a}_{2}$, ${\bf a}_{3}$ are a set of primitive
translation vectors for the lattice \cite{kittel}.
We choose the orientation of the ${\bf a}_{i}$ in such a way
that $V={\bf a}_{1}\cdot {\bf a}_{2}\times {\bf a}_{3}>0$. $V$ clearly
represents the volume of the primitive cell.
Each point ${\bf r}_{{\bf n}}$
is the equilibrium position of a point particle (electron)
of mass $m$ and
electric charge $e$, which is subject to an elastic force of the form
\[
{\bf F}_{{\rm el}}=-K({\bf z}_{{\bf n}}-{\bf r}_{{\bf n}})\,,
\]
where the vector ${\bf z}_{{\bf n}}$ represents the instantaneous
coordinates of the point particle. Although it is not strictly
necessary for the mathematical self-consistency of the model, in
order to reproduce in a more realistic way the situation of solid
state physics we can assume that ${\bf r}_{{\bf n}}$\ is also the
seat of a static positive ion of charge $-e$. As is traditionally
the case for the classical models of dispersion, we shall neglect
the Coulomb interaction between the ion and its associated
electron, since classical mechanics fails at such short distance
scales, and we shall instead identify the proper oscillator
frequency $\omega _{0}=\sqrt{K/m}$ with a characteristic
excitation frequency of the optical electron in the atomic ground
state. According to quantum mechanics, it could even be possible
to obtain a more realistic model by associating to each atom a set
of fictitious oscillators \cite{becker}, one for each allowed
quantum transition from the ground state of energy $E_{0}$ to an
excited state of energy $E_{n}$, with proper frequencies $\omega
_{0n}=(E_{n}-E_{0})/\hbar $. In order to assign to the
contribution of each oscillator an appropriate weight, one has
then to make the substitution $e^{2}/m\rightarrow f_{0n}e^{2}/m$,
where the ``oscillator strengths'' $f_{0n}$\ are coefficients
subject to the sum rule $\sum_{n}f_{0n}=1$ (we are considering
here atoms with a single optically active electron). An explicit
calculation from first order perturbation theory provides
\[
f_{0n}=(2m/\hbar )\omega _{0n}\left| \left\langle E_{0}\left|
\bm {\varepsilon} \cdot {\bf \hat{x}}\right| E_{n}
\right\rangle \right|^{2},
\]
${\bf \hat{x}}$ being the position operator and $\bm
{\varepsilon}$ the unit vector representing the direction of
oscillation. For the sake of simplicity we shall consider in the
following calculations a single oscillator per atom, in accordance
with the original literature on classical dispersion, although the
extension to the case of multiple oscillators presents no
conceptual difficulty.

The charge and current densities associated with the electron-ion pair
are given respectively by
\begin{eqnarray}
\rho _{{\bf n}}({\bf x},t)&=& e\left[ \delta ^{3}({\bf x}-{\bf
z}_{{\bf n}}(t))
-\delta ^{3}({\bf x}-{\bf r}_{{\bf n}})\right]  \label{rho} \\
{\bf j}_{{\bf n}}({\bf x},t)&=& e{\bf \dot{z}}_{{\bf n}}(t)
\delta ^{3}({\bf x}-{\bf z}_{{\bf n}}(t))\,,  \label{jey}
\end{eqnarray}
where $\delta $ denotes Dirac's delta function. The retarded
potentials generated by the charge-current density $j_{{\bf
n}}^{\mu }$ (with $j^0_{\bf n}\equiv c\rho_{\bf n}$) are defined
as
\begin{eqnarray}
A_{{\bf n}}^{\mu,{\rm ret}}(x)&=& \frac{1}{c}\int d^{4}y\,D_{{\rm
ret}}(x-y) j_{{\bf n}}^{\mu }(y) \nonumber \\
&=& \frac{1}{(2\pi )^{4}c} \int d^{4}k\,\frac{e^{ik\cdot
x}}{k^{2}-i\varepsilon k^{0}} \tilde{j}_{{\bf n}}^{\mu }(k)\,,
\label{potential}
\end{eqnarray}
where
\begin{equation}
\,D_{{\rm ret}}(x)=\frac{\delta (x^{0}-\left| {\bf x}\right| )}{4\pi \left|
{\bf x}\right| }=\frac{1}{(2\pi )^{4}}\int d^{4}k\,\frac{e^{ik\cdot x}}
{k^{2}-i\varepsilon k^{0}}  \label{green}
\end{equation}
is the retarded Green function \cite{jackson} and
\begin{equation}
\tilde{j}_{{\bf n}}^{\mu }(k)=\int d^{4}y\,e^{-ik\cdot y}
j_{{\bf n}}^{\mu}(y)  \label{jtilde}
\end{equation}
is the Fourier transform of $j_{{\bf n}}^{\mu }$.
We are here using the
four-dimensional notation so that, for instance, $x$ denotes the
four-vector $(x^{0},{\bf x})$, and $x^{0}=ct$,
$k\cdot x=k_{\mu }x^{\mu }={\bf k}%
\cdot {\bf x}-k^{0}x^{0}$, $k^{2}=k\cdot k={\bf k}^{2}-(k^{0})^{2}$.
Summation over repeated indices is always implicitly understood. The
retarded potentials satisfy the equation
\[
\partial ^{\nu }\partial _{\nu }A_{{\bf n}}^{\mu,{\rm ret}}
=-\frac{1}{c}j_{{\bf n}}^{\mu }
\]
and the Lorentz gauge condition
\[
\partial _{\mu }A_{{\bf n}}^{\mu,{\rm ret}}=0\,.
\]
Introducing then the retarded fields according to the usual relations
\begin{eqnarray}
{\bf E}_{\bf n}^{\rm ret}&=&-{\bf \nabla }A_{\bf n}^{0,{\rm ret}}
-\frac{\partial }{c\partial t}
{\bf A}_{\bf n}^{\rm ret}  \label{field} \\
{\bf B}_{\bf n}^{\rm ret}&=&{\bf \nabla}\times
{\bf A}_{\bf n}^{\rm ret}
\end{eqnarray}
and putting ${\bf x}_{{\bf n}}={\bf z}_{{\bf n}}-{\bf r}_{{\bf n}}$,
we can write the (nonrelativistic)
equation of motion of the electrons as
\begin{eqnarray}
m{\bf \ddot{x}}_{{\bf n}}&=&-K{\bf x}_{{\bf n}}+e {\bf e}_{{\bf
n}} ({\bf r}_{{\bf n}}+{\bf x}_{{\bf n}},t) \nonumber \\
&&+ e{\bf \dot{x}}_{{\bf n}}\times {\bf b}_{{\bf n}}({\bf r}_{{\bf
n}}+{\bf x}_{{\bf n}},t) +\frac{e^{2}}{6\pi c^{3}} \dddot{\bf
x}_{{\bf n}}\,, \label{eqmoto}
\end{eqnarray}
where
\begin{eqnarray}
{\bf e}_{{\bf n}}&=&\sum_{{\bf m}\neq {\bf 0}}{\bf E}_{{\bf n}+{\bf m}}
^{\rm ret}  \label{erregende} \\
{\bf b}_{{\bf n}}&=&\sum_{{\bf m}\neq {\bf 0}}{\bf B}_{{\bf n}+{\bf m}}
^{\rm ret}
\end{eqnarray}
represent the exciting fields. The notation used in the two last
equations means that the summation index $\bf m$ runs over all the
values in $\mathbb{Z}^{3}$ except $\bf 0$. The last term in Eq.\
(\ref{eqmoto}) describes the radiation reaction force, according
to the Lorentz--Dirac prescription. See Ref.\ \cite{milonni} for a
comparison between the expression of the atomic polarizability
resulting from this classical equation in the dipole
approximation, and the corresponding result obtained for a
two-level atom in electric-dipole interaction with the quantized
electromagnetic field.

\section{Normal-mode solutions}

We suppose the amplitude of the oscillations to be small enough,
so that at every stage we can neglect all terms of order higher
than one in the ${\bf x}_{{\bf n}}$ and their time derivatives of
any order. It follows that Eq.\ (\ref{eqmoto}) simplifies to
\begin{equation}
m{\bf \ddot{x}}_{{\bf n}}=-K{\bf x}_{{\bf n}}+e{\bf e}_{{\bf n}}
({\bf r}_{{\bf n}},t)+\frac{e^{2}}{6\pi c^{3}}
\dddot{\bf x}_{{\bf n}}\,, \label{eqmotod}
\end{equation}
where the retarded fields, included into ${\bf e}_{{\bf n}}$ according to
Eq.\ (\ref{erregende}), are to be calculated in the dipole approximation,
whereby each ${\bf E}_{\bf m}^{\rm ret}$
depends linearly on ${\bf x}_{{\bf m}}$ and its time derivatives.
Of course, Eq.\ (\ref{eqmotod}) actually
represents an infinite system of coupled linear equations,
since we have one
such equation for each ${\bf n}\in \mathbb{Z}^{3}$.
We shall look for
a global solution of the form (in customary complex notation)
\begin{equation}
{\bf x}_{{\bf n}}={\bf C}\exp \left[ i(\bm{\kappa} \cdot
{\bf r}_{{\bf n}}-\omega t)\right]\,,  \label{normmode}
\end{equation}
representing a plane wave with amplitude ${\bf C}$, frequency
$\omega $ and wavevector $\bm{\kappa}$. The parameters $\omega $
and $\bm{\kappa}$\ can always be chosen so that $\omega >0$\ and
$\bm{\kappa}$\ belongs to the first Brillouin zone of the crystal.
We shall proceed as follows: for a generic motion of the form
(\ref{normmode}) we shall calculate the resulting expression for
the exciting field ${\bf e}_{{\bf n}}$; then by substituting this
expression into Eq.\ (\ref{eqmotod}) we shall find out that the
equations of motion of
all the particles can be simultaneously satisfied, provided that
$\bm{\kappa}$\ and $\omega $ satisfy a well defined dispersion relation.

In the dipole approximation, that is
to first order in ${\bf C}$, the charge and current densities become
\begin{eqnarray*}
\rho _{{\bf n}}({\bf x},t) &=&-e
{\bf C\cdot \nabla }\delta ^{3}({\bf x}-{\bf r}_{{\bf n}})
\exp \left[ i(\bm{\kappa}
\cdot {\bf r}_{{\bf n}}-\omega t)\right]  \\
{\bf j}_{{\bf n}}({\bf x},t) &=&-ie\omega {\bf C} \delta ^{3}({\bf
x}-{\bf r}_{{\bf n}})\exp \left[ i(\bm{\kappa} \cdot {\bf r}_{{\bf
n}}-\omega t)\right]
\end{eqnarray*}
whence, according to the definition (\ref{jtilde})
\begin{eqnarray}
\tilde{\rho}_{{\bf n}}(k) &=&-2\pi ie
{\bf k\cdot C}\delta (k^{0}-\omega
/c)\exp \left[ i(\bm{\kappa}-{\bf k}){\bf \cdot r}_{{\bf n}}\right]
\label{rhojtilde} \\
{\bf \tilde{j}}_{{\bf n}}(k) &=&-2\pi ie
\omega {\bf C}\delta (k^{0}-\omega
/c)\exp \left[ i(\bm{\kappa}-{\bf k}){\bf \cdot r}_{{\bf n}}
\right]\,.  \label{jeytilde}
\end{eqnarray}
Substituting these expressions into Eq.\ (\ref{potential}) and
then applying Eq.\ (\ref{field}) we get
\begin{eqnarray}
{\bf E}_{\bf m}^{\rm ret}({\bf x},t)&=&e\exp \left[ i(\bm{\kappa}
\cdot {\bf r}_{{\bf m}}-\omega t)\right] \nonumber \\
&&\times \int \frac{d^{3}{\bf k}}{(2\pi )^{3}}\,\frac {(\omega
/c)^{2}{\bf C}-({\bf k}\cdot {\bf C}){\bf k}} {{\bf
k}^{2}-(\omega/c)^{2}-i\varepsilon }\nonumber \\
&&\times\exp \left[ i{\bf k\cdot (x}-{\bf r}_{{\bf m}})\right]\,.
\label{eenne}
\end{eqnarray}
It follows
\begin{equation}
{\bf e}_{{\bf n}}({\bf r}_{{\bf n}},t)=e{\bf \hat{L}}^{{\rm ret}}
(\bm{\kappa},\omega )\cdot {\bf C}
\exp \left[ i(\bm{\kappa}\cdot
{\bf r}_{{\bf n}}-\omega t)\right] \,,  \label{erreg1}
\end{equation}
where the second rank tensor ${\bf \hat{L}}^{{\rm ret}}$ has components
\begin{eqnarray*}
L^{{\rm ret}}_{ij}(\bm{\kappa},\omega )&=& \sum_{{\bf m}\neq {\bf
0}}\int \frac{d^{3}{\bf k}}{(2\pi )^{3}}\,\frac{(\omega
/c)^{2}\delta _{ij}- k_{i}k_{j}}{{\bf k}^{2}-(\omega
/c)^{2}-i\varepsilon } \\
&&\times\exp \left[ i({\bf k}-\bm{\kappa} )\cdot {\bf r}_{{\bf
m}}\right]
\end{eqnarray*}
and we have used the notation ${\bf \hat{L}}^{{\rm ret}}\cdot {\bf
C}$ to denote the vector with components $L^{{\rm
ret}}_{ij}C_{j}$. By substituting the expressions (\ref{normmode})
and (\ref{erreg1}) into Eq.\ (\ref{eqmotod}), the original system
of infinitely many coupled equations is transformed into the
single vectorial equation
\begin{equation}
\frac{e^{2}}{m}{\bf \hat{L}}^{{\rm ret}}(\bm{\kappa},\omega )\cdot {\bf C}
=\left( \omega _{0}^{2}-\omega ^{2}-i\frac{\omega ^{3}e^{2}}{6\pi mc^{3}}
\right) {\bf C}\,,  \label{eqmotod1}
\end{equation}
which admits solutions with nonvanishing ${\bf C}$\ when
\begin{equation}
\det \left[ \frac{e^{2}}{m}{\bf \hat{L}}^{{\rm ret}}(\bm{\kappa},
\omega)-\left( \omega _{0}^{2}-\omega ^{2}-i\frac{\omega ^{3}e^{2}}
{6\pi mc^{3}}\right) {\bf \hat{1}}\right] =0\,.  \label{determ}
\end{equation}
The above equation determines an implicit relation between
$\bm{\kappa}$ and $\omega $,
which constitutes the sought for dispersion relation of the crystal.

Once a solution of the form (\ref{normmode}) has been obtained, it
is easy to write down the expression for the macroscopic
quantities which can be associated to the normal mode. The
macroscopic polarization density is in fact given by the
continuous function of space which interpolates the microscopic
displacement vectors of the individual dipoles:
\begin{equation}
{\bf P} ({\bf x},t)=V^{-1}e{\bf C}\exp \left[
i(\bm{\kappa} \cdot {\bf x}-\omega t)\right]\,. \label{pmacr}
\end{equation}
The macroscopic fields ${\bf E}_{\rm mac}$, ${\bf B}_{\rm mac}$
and ${\bf D}_{\rm mac}\equiv
{\bf E}_{\rm mac}+{\bf P}$ are then given by the solutions
of the usual macroscopic Maxwell equations
inside the (nonmagnetic) medium:
\begin{eqnarray}
{\bf \nabla} \cdot {\bf D}_{\rm mac}&=&0 \label{maxwell1}\\
{\bf \nabla} \times {\bf E}_{\rm mac} &=&-\frac{\partial}
{c\partial t} {\bf B}_{\rm mac} \label{maxwell2} \\
{\bf \nabla} \cdot {\bf B}_{\rm mac} &=&0 \\
{\bf \nabla} \times {\bf B}_{\rm mac} &=&\frac{\partial}
{c\partial t}{\bf D}_{\rm mac}\,. \label{maxwell4}
\end{eqnarray}
From Eq.\ (\ref {maxwell1}) it follows ${\bf \nabla} \cdot {\bf
E}_{\rm mac} =-{\bf \nabla} \cdot {\bf P}$. Then by
taking the curl of Eq.\ (\ref {maxwell2}) and eliminating ${\bf
B}_{\rm mac}$ with the aid of Eq.\ (\ref{maxwell4}) one easily
obtains
\[
\Delta {\bf E}_{\rm mac} -\frac{\partial^2}{c^2 \partial t^2}
{\bf E}_{\rm mac} =\frac{\partial^2}{c^2\partial t^2} {\bf P}
-{\bf \nabla}({\bf \nabla}\cdot {\bf P})\,.
\]
From this, Eq.\ (\ref {pmacr}) and Eq.\ (\ref {maxwell2}), we conclude
\begin{eqnarray}
{\bf E}_{\rm mac}({\bf x},t)&=&\frac{e}{V}\frac{(\omega /c)^2{\bf
C} -\bm{\kappa}(\bm{\kappa}\cdot{\bf C})}{\bm{\kappa}^2-(\omega
/c)^2} \nonumber \\
&&\times\exp \left[ i(\bm{\kappa}\cdot {\bf x}-\omega t)\right]
\label{emacro} \\
{\bf B}_{\rm mac}({\bf x},t)&=&\frac{e}{V}\frac{(\omega /c)
\bm{\kappa} \times {\bf C}}{\bm{\kappa}^2-(\omega /c)^2} \nonumber
\\
&&\times\exp \left[ i(\bm{\kappa}\cdot {\bf x}-\omega t)\right]\,.
\end{eqnarray}

\section{The Wheeler--Feynman identity}

Since (\ref{determ}) is a complex equation, it is
{\it a priori} to be expected that, in order that a
real solution for $\omega $ may exist, the vector $\bm{\kappa}$
must necessarily be assigned an imaginary component,
which represents an extinction coefficient for the wave. A
fundamental observation can however be made at this point, showing
that this is not actually the case and that Eq.\ (\ref{determ})
determines $\omega $\ as a real function of the real independent
variable $\bm{\kappa}$. To this purpose, let us introduce the
advanced potentials $A_{\bf n}^{\mu,{\rm adv}}$
defined by a formula analogous to (\ref{potential}) with,
in place of $D_{{\rm ret}}(x)$, the advanced Green function
\[
D_{{\rm adv}}(x)=\frac{\delta (x^{0}+\left| {\bf x}\right| )}{4\pi \left|
{\bf x}\right| }=\frac{1}{(2\pi )^{4}}\int d^{4}k\,\frac{e^{ik\cdot x}}
{k^{2}+i\varepsilon k^{0}}\,.
\]
A completely general result about the Lorentz--Dirac equation
asserts that the self-force, given by the expression involving the
triple time-derivative of the particle position, is equal to the
electromagnetic force exerted on the particle by one half the
difference between the retarded and advanced fields generated by
the particle itself \cite{dirac,marino}. Using this result in the
dipole approximation, we have that the last term of Eq.\
(\ref{eqmotod}) can be expressed as
\begin{equation}
\frac{e^{2}}{6\pi c^{3}} \dddot{\bf x}_{{\bf n}}=e{\bf E}_{{\bf n}}^{(-)}
({\bf r}_{{\bf n}})\,,  \label{dirac}
\end{equation}
where $A_{{\bf n}}^{\mu(\pm )}\equiv (A_{\bf n}^{\mu,{\rm ret}}\pm
A_{\bf n}^{\mu,{\rm adv}})/2$ and
${\bf E}_{\bf n}^{\pm}=-{\bf \nabla }A_{\bf n}^{0,\pm}
-(\partial /c\partial t){\bf A}_{\bf n}^{\pm}$.
Note that the field ${\bf E}_{{\bf n}}^{(-)}$\
is a solution of the homogeneous (i.e.\ source-free) field equation and
therefore, at variance with ${\bf E}_{{\bf n}}^{(+)}$,
it is regular at the particle position ${\bf r}_{{\bf n}}$.
Using Eq.\ (\ref{dirac}) and the identity ${\bf E}_{\bf n}^{\rm ret}=
{\bf E}_{{\bf n}}^{(+)}+{\bf E}_{{\bf n}}^{(-)}$,
we can rewrite Eq.\ (\ref{eqmotod}) as
\begin{equation}
m{\bf \ddot{x}}_{{\bf n}}=-K{\bf x}_{{\bf n}}+e{\bf e}_{{\bf n}}^{(+)}
({\bf r}_{{\bf n}},t)
+e{\bf E}^{(-)}({\bf r}_{{\bf n}},t)\,,  \label{eqmotod2}
\end{equation}
where
\begin{eqnarray*}
{\bf e}_{{\bf n}}^{(\pm)} &=&\sum_{{\bf m}\neq {\bf 0}}
{\bf E}_{{\bf n}+{\bf m}}^{(\pm)}
\\
{\bf E}^{(-)} &=&\sum_{{\bf m}}{\bf E}_{{\bf m}}^{(-)}\,,
\end{eqnarray*}
the summation being extended to all the values
${\bf m}\in \mathbb{Z}^{3}$ in the last equation. We have
\begin{eqnarray}
A^{\mu(-)}(x)&=&\sum_{\bf m}A_{\bf m}^{\mu(-)}(x) \nonumber
\\
&=&\sum_{\bf m}
\frac{1}{c}\int d^{4}y\,D^{(-)}(x-y)j^{\mu}_{\bf m}(y)  \nonumber \\
&=&\frac{i}{2(2\pi )^{3}c}
\int d^{4}k\,e^{ik\cdot x}\delta (k^{2})
\varepsilon (k^{0}) \tilde{j}^{\mu }(k)\,,  \label{ameno}
\end{eqnarray}
with
\begin{eqnarray*}
D^{(-)}(x)&\equiv& \frac{1}{2}\left[ D_{{\rm ret}}(x)-D_{{\rm
adv}}(x) \right] \nonumber \\
&=&\frac{\delta (x^{2}) \varepsilon (x^{0})}{4\pi }=
\frac{i}{2(2\pi )^{3}}\int d^{4}k\,e^{ik\cdot x} \delta
(k^{2})\varepsilon (k^{0})
\end{eqnarray*}
and
\[
\tilde{j}^{\mu }(k)=\sum_{{\bf m}}
\tilde{j}_{{\bf m}}^{\mu}(k)\,.
\]
For a normal mode solution the above threefold series can be evaluated
by using Eqs.\ (\ref{rhojtilde}-\ref{jeytilde}) and the relation
\begin{equation}
\sum_{{\bf l}}\exp (i{\bf k\cdot r}_{{\bf l}})=\frac{(2\pi
)^{3}}{V}\sum_{{\bf m}}\delta ^{3}
({\bf k}-{\bf G}_{{\bf m}})\,,  \label{recipro}
\end{equation}
where the ${\bf G}_{{\bf m}}$, for ${\bf m}\in \mathbb{Z}^{3}$,
are the points of the reciprocal lattice \cite{kittel}, defined as
\[
{\bf G}_{{\bf m}}=\frac{\pi }{V}\varepsilon _{ijk}m_{i}
{\bf a}_{j}\times {\bf a}_{k}\,.
\]
Here $\varepsilon _{ijk}$\ indicates the completely antisymmetric tensor
with $\varepsilon _{123}=1$. We obtain
\begin{eqnarray}
\tilde{\rho}(k) &=&-ie{\bf k\cdot C}\frac{(2\pi )^{4}}{V}\delta
(k^{0}- \omega/c) \nonumber \\
&&\times\sum_{{\bf m}}\delta ^{3}({\bf k}-\bm{\kappa}-
{\bf G}_{{\bf m}}) \\
{\bf \tilde{j}}(k) &=&-ie\omega {\bf C}\frac{(2\pi )^{4}}{V}\delta
(k^{0}-\omega /c) \nonumber \\
&&\times\sum_{{\bf m}}\delta ^{3}({\bf k}- \bm {\kappa }-{\bf
G}_{{\bf m}})\,.
\end{eqnarray}
We see that the integrand on the r.h.s.\ of Eq.\ (\ref{ameno}) is
a singular function with support on the cone $k^{2}=|{\bf
k}|^{2}-(k^{0})^{2}=0$. On the other hand, for a given real
$\bm{\kappa}$, the functions $\tilde{\rho}(k)$\ and ${\bf
\tilde{j}}(k)$\ are different from zero on this cone only when
there exists ${\bf m}\in \mathbb{Z}^{3}$\ such that $\omega =
c|\bm{\kappa}+{\bf G}_{{\bf m}}| \equiv \omega _{{\bf
m}}(\bm{\kappa})$. This proves that for all values of $\omega $,
except those belonging to the discrete set of singular values
$\left\{\omega_{{\bf m}}(\bm{\kappa}) \right\} _{{\bf m}\in
\mathbb{Z}^{3}}$, the ``Wheeler--Feynman identity''
\[
A_{\mu }^{(-)}(x)=0
\]
holds at any point $x$\ of spacetime. On the other hand, the
results that will be obtained in the next section show immediately
that, for a given $\bm{\kappa}$, the exciting field diverges for
$\omega = \omega _{{\bf m}}(\bm{\kappa})$, so that none of these
frequency values can possibly correspond to a normal mode
solution. The Wheeler--Feynman identity is therefore established
in complete generality for all physical solutions expressible as
linear combinations of normal modes. As an immediate consequence
of this identity one has that the last term on the r.h.s.\ of Eq.\
(\ref{eqmotod2}) vanishes. Recalling Eq.\ (\ref{dirac}), this
result can also be put in the form
\[
{\bf e}_{{\bf n}}^{(-)}({\bf r}_{{\bf n}},t)=
-\frac{e}{6\pi c^{3}} \dddot{\bf x}_{{\bf n}}\,.
\]
A relation physically equivalent to this one was obtained by Oseen
(although without a rigorous mathematical proof) already in 1916
\cite{oseen1}.

We can then write
\[
{\bf e}_{{\bf n}}^{(\pm)}({\bf r}_{{\bf n}},t)=
e{\bf \hat{L}}^{(\pm)}(\bm {\kappa},\omega )\cdot {\bf C}\exp
\left[ i(\bm {\kappa} \cdot {\bf r}_{{\bf n}}-\omega t)\right]\,,
\]
with
\begin{eqnarray}
L^{(-)}_{ij}(\bm{\kappa},\omega )&=&
-\frac{i}{6\pi}\left(\frac{\omega}{c}\right)^3\delta_{ij}  \label{oseen} \\
L^{(+)}_{ij}(\bm{\kappa},\omega )&=&\sum_{{\bf m}\neq {\bf
0}}P\int \frac{d^{3}{\bf k}}{(2\pi )^{3}}\,\frac {(\omega
/c)^{2}\delta _{ij}-k_{i}k_{j}}{{\bf k}^{2}-(\omega /c)^{2}}
\nonumber \\
&&\times\exp \left[ i({\bf k}-\bm {\kappa})\cdot {\bf r}_{{\bf m}}
\right] \,.
\end{eqnarray}
The symbol $P\int $ indicates the principal value of the integral.
Note that for real $\bm{\kappa}$ and $\omega $\ the function
$L^{(+)}_{ij}(\bm {\kappa },\omega )$\ is real, since taking the
complex conjugate amounts to making the substitution ${\bf m}$
$\rightarrow -{\bf m}$\ in the summation index. From these
considerations it follows that the equation of motion (\ref
{eqmotod1}) can be rewritten as
\begin{equation}
\frac{e^{2}}{m}{\bf \hat{L}}^{(+)}(\bm{\kappa},\omega )\cdot {\bf C}
=\left( \omega _{0}^{2}-\omega ^{2}\right) {\bf C}\,,  \label{eqmotoel}
\end{equation}
and the original complex equation (\ref{determ}) is converted
into the real equation
\[
\det \left[ \frac{e^{2}}{m}{\bf \hat{L}}^{(+)}(\bm{\kappa},\omega )-
\left(\omega _{0}^{2}-\omega ^{2}\right) {\bf \hat{1}}\right] =0\,,
\]
which determines a dispersion relation between the real variables
$\bm {\kappa }$ and $\omega $. Note that this remarkable result,
which allows for the propagation of undamped plane waves in the
crystal, holds just as a consequence of the inclusion of the
Lorentz--Dirac radiation reaction term in the equations of motion.
This does not appear surprising, when one recalls that the
expression of this term was determined just in order to insure
global energy conservation for the complete system of particles
and field \cite{dirac,marino}.

\section{The calculation of the exciting field}

\subsection{Outline of the procedure}

The expression for the retarded field produced by an oscillating
dipole is well-known, and it can in fact be obtained by explicit
calculation of the integral on the r.h.s.\ of Eq.\ (\ref{eenne}).
It might seem therefore that the most direct way of calculating
the exciting field ${\bf e}_{{\bf n}}$ would be to substitute such
an expression into Eq.\ (\ref{erregende}), as was done in Ref.\
\cite{CG} for the one-dimensional case. It turns out however that
in the three-dimensional case the series on the r.h.s.\ of Eq.\
(\ref{erregende}) does not converge in a proper sense. It is
possible indeed to assign to the sum an unambiguous meaning via
the prescription
\[
{\bf e}_{{\bf n}}({\bf r}_{{\bf n}},t)=
\lim_{\eta \rightarrow 0^{+}}\sum_{{\bf m}\neq {\bf 0}}
{\bf E}_{{\bf n}+{\bf m}}^{\rm ret}({\bf r}_{{\bf n}},t)\exp
\left( -\eta {\bf r}_{{\bf m}}^{2}\right)\,,
\]
but the above formula is not convenient
for numerical computations purposes
(the sum converges slowly for nearly vanishing $\eta $),
and furthermore
gives us no insight into the physical content of the results.
We shall therefore follow a different path,
which consists in converting the sum over
the points ${\bf r}_{{\bf n}}$ into a sum over the points
${\bf G}_{{\bf m}}$\ of the reciprocal lattice.
After some manipulations
we shall obtain an absolutely convergent
series, which in typical cases can be
numerically computed with little
effort. Furthermore, our final expression
will provide an explicit expansion
in powers of $a/\lambda $ ($a=V^{1/3}$ being the mean lattice parameter
and $\lambda =2\pi /\kappa $ the
wavelength), in such a way that the well-known
result provided by the old theories in the long-wavelength
limit will appear
to be just the zero-order approximation of the general result.

We start from the relation
\begin{equation}
{\bf e}_{{\bf n}}({\bf r}_{{\bf n}},t)=
\lim_{{\bf x}\rightarrow {\bf r}_{{\bf n}}}
\left[ {\bf E}^{{\rm ret}}({\bf x},t)-{\bf E}_{\bf n}^{\rm ret}
({\bf x},t)\right] \,,  \label{erreg2}
\end{equation}
where
\[
{\bf E}^{{\rm ret}}({\bf x},t)=\sum_{{\bf l}}
{\bf E}_{\bf l}^{\rm ret}({\bf x},t)
\]
is the total retarded field. Using Eqs.\ (\ref{eenne}) and
(\ref{recipro}) we obtain
\begin{eqnarray}
{\bf E}^{{\rm ret}}({\bf x},t)&=&\frac eV \sum_{{\bf m}}
\frac{(\omega /c)^{2}{\bf C}- \left[ ({\bf G}_{{\bf
m}}+\bm{\kappa})\cdot {\bf C}\right] ({\bf G}_{{\bf
m}}+\bm{\kappa})}{({\bf G}_{{\bf m}}+ \bm{\kappa})^{2}-(\omega
/c)^{2}-i\varepsilon } \nonumber \\
&&\times\exp \left[ i({\bf G}_{{\bf m}}+\bm {\kappa })\cdot {\bf
x} -i\omega t\right] \,. \label{etot}
\end{eqnarray}
Note that, according to Eq.\ (\ref{emacro}), the term for ${\bf
m}=0$ of the above series is equal to the macroscopic field ${\bf
E}_{\rm mac}$. This of course corresponds to the fact that, for
$\lambda \gg a$, the macroscopic field just represents the average
over a unit lattice cell of the total microscopic retarded field.
We observe also that ${\bf r}_{{\bf n}}\cdot {\bf G}_{{\bf
m}}=2\pi {\bf n\cdot m}$, so that
\begin{eqnarray*}
&&\exp \left[ i({\bf G}_{{\bf m}}+ \bm{\kappa})\cdot {\bf x}
\right] \\
&=&\exp [i\bm {\kappa} \cdot {\bf r}_{{\bf n}}+i({\bf G}_{{\bf m}}
+\bm{\kappa})\cdot ({\bf x}-{\bf r}_{{\bf n}})]\,.
\end{eqnarray*}
Therefore, substituting Eqs.\ (\ref{etot}) and (\ref{eenne}) into
Eq.\ (\ref{erreg2}), and then comparing the resulting expression
with Eq.\ (\ref{erreg1}), we get
\begin{eqnarray*}
VL^{{\rm ret}}_{ij}(\bm{\kappa},\omega )&=&\lim_{{\bf
x}\rightarrow {\bf 0}} \int d^{3}{\bf p}\,e^{i{\bf p\cdot
x}}\frac{f^{2}\delta _{ij} -p_{i}p_{j}}{{\bf
p}^{2}-f^{2}-i\varepsilon } \\
&&\times\left[ \sum_{{\bf m}}\delta ^{3} ({\bf p-H}_{{\bf m}}-{\bf
q})-1\right] \equiv M^{{\rm ret}}_{ij}({\bf q},f)\,,
\end{eqnarray*}
where we have introduced the dimensionless variables $f=\omega
a/2\pi c$, ${\bf q}=\bm{\kappa}a/2\pi $, and ${\bf H}_{{\bf m}}
=a{\bf G}_{{\bf m}}/2\pi $. Let us now shortly rewrite the r.h.s.\
of the above equation as $\lim_{{\bf x}\rightarrow {\bf 0}}F({\bf
x})$, with $F({\bf x})=\int d^{3}{\bf p}\,e^{i{\bf p\cdot
x}}\tilde{F}({\bf p})$. The regularity of the function $F({\bf
x})$ at ${\bf x}={\bf 0}$ descends obviously from the fact that
${\bf e}_{{\bf n}}({\bf x},t)$ is regular at ${\bf x}={\bf
r}_{{\bf n}}$, as can be seen from the very definition
(\ref{erregende}) of ${\bf e}_{{\bf n}}$. Recalling that
\begin{eqnarray}
\frac{\exp (-{\bf x}^{2}/4\eta )}{8(\pi \eta )^{3/2}} &=&\int
\frac{d^{3}{\bf k}}{(2\pi )^{3}} \exp (-\eta {\bf k}^{2}-i{\bf
k}\cdot {\bf x}) \nonumber \\
&\rightarrow& \delta ^{3}({\bf x})\qquad \text{for }\eta
\rightarrow 0^{+},  \label{gauss}
\end{eqnarray}
we can write
\begin{eqnarray*}
\lim_{{\bf x}\rightarrow {\bf 0}}F({\bf x}) &=&\lim_{\eta \rightarrow
0^{+}}\int d^{3}{\bf x}\,F({\bf x})\frac{\exp(-{\bf x}^{2}/4\eta
-i{\bf q}\cdot {\bf x})}{8(\pi \eta )^{3/2}} \\
&=&\lim_{\eta \rightarrow 0^{+}}\int d^{3} {\bf x}\int d^{3}{\bf
p}\,e^{i({\bf p}-{\bf q})\cdot {\bf x}} \tilde{F}({\bf p}) \\
&&\times\int \frac{d^{3}{\bf k}}{(2\pi )^{3}}
e^{-\eta {\bf k}^{2}-i{\bf k}\cdot {\bf x}} \\
&=&\lim_{\eta \rightarrow 0^{+}}\int d^{3}{\bf k}\,\tilde{F}
({\bf k}+{\bf q})e^{-\eta {\bf k}^{2}}.
\end{eqnarray*}
Reintroducing the explicit expression for $\tilde{F}$, we thus obtain
\begin{eqnarray}
M^{{\rm ret}}_{ij}({\bf q},f) &=&-\lim_{\eta \rightarrow 0^{+}}
\int d^{3}{\bf k}\,c_{ij}({\bf k},{\bf q},f) \nonumber \\
&&\times\left[ \sum_{{\bf m}}\delta ^{3}({\bf k}-{\bf H}_{{\bf
m}})-1\right] e^{-\eta {\bf k}^{2}} \label{mret0}
\end{eqnarray}
with
\[
c_{ij}({\bf k},{\bf q},f) =\frac{(k_{i}+q_{i})(k_{j}+q_{j})
-f^{2}\delta_{ij}}{({\bf k}+{\bf q})^{2}-f^{2}-i\varepsilon }\,.
\]

By isolating the term for ${\bf m}={\bf 0}$ of
the series on the r.h.s.\
of Eq.\ (\ref{mret0}) we get the expression
\begin{equation}
-c_{ij}({\bf 0},{\bf q},f) =
\frac{f^{2}\delta _{ij}-q_{i}q_{j}}{{\bf q}^{2}-f^{2}} \equiv
M^{{\rm mac}}_{ij}({\bf q},f) \,,
\end{equation}
which is related to the macroscopic field inside the crystal
according to Eq.\ (\ref{emacro}):
\begin{equation}
{\bf E}_{\rm mac}({\bf x},t)= eV^{-1}{\bf \hat{M}}^{{\rm mac}}
({\bf q},f)\cdot {\bf C}\exp
\left[ i(\bm{\kappa}\cdot{\bf x}-\omega t)\right]\,.
\label{emacro2}
\end{equation}
We can then write
\begin{eqnarray}
M^{{\rm ret}}_{ij}({\bf q},f) &=& M^{{\rm mac}}_{ij}({\bf q},f)
\nonumber \\
&& -\lim_{\eta \rightarrow 0^{+}}\left[ S_{ij}({\bf q},f,\eta
)-I_{ij}({\bf q},f,\eta )\right] \,,
\end{eqnarray}
with
\begin{eqnarray*}
S_{ij}(\eta ,{\bf q},f) &=&\sum_{{\bf m}\neq {\bf 0}}c_{ij}
({\bf H}_{{\bf m}},{\bf q},f)
\exp \left( -\eta {\bf H}_{{\bf m}}^{2}\right) \\
I_{ij}(\eta ,{\bf q},f) &=&\int d^{3}{\bf k}\,c_{ij}({\bf k},{\bf q},f)
\exp (-\eta {\bf k}^{2})\,.
\end{eqnarray*}
The series $S_{ij}$ and the integral $I_{ij}$ are both divergent
in the limit $\eta \rightarrow 0$.
We are going to study them separately and to
split each of them into a divergent and a convergent part.
The two divergent
parts must of course cancel each other, as we shall check directly.
We shall then be able to express $M^{{\rm ret}}_{ij}$ as the
sum of a finite term and of
an absolutely convergent series.

By expanding the function $c_{ij}$ in powers of $1/k$ (from now
on $k$ stands for $|{\bf k}|$) we can write
\[
c_{ij}=c_{ij}^{(0)}+c_{ij}^{(1)}+c_{ij}^{(2)}+c_{ij}^{(3)}\theta
(k-\varepsilon )+\bar{c}_{ij},
\]
where, for $l=0,\ldots,3$, $c_{ij}^{(l)}$ is a
homogeneous function of ${\bf k}$ of degree $-l$,
and $\bar{c}_{ij}=O(k^{-4})$ for $k\rightarrow
\infty $. The term of degree $-3$ has been multiplied by $\theta
(k-\varepsilon )$\ [where $\theta (x)=1$\ for $x\geq 0$,
$\theta (x)=0$\ for $x<0$)]
in order that all terms appearing in the above equation be
integrable in a neighborhood of ${\bf k}=0$. We have
\begin{eqnarray*}
c_{ij}^{(0)} &=&\frac{k_{i}k_{j}}{k^{2}} \\
c_{ij}^{(1)} &=&\frac{1}{k^{2}}\left( q_{i}k_{j}+q_{j}k_{i}-2{\bf q}\cdot
{\bf k}\frac{k_{i}k_{j}}{k^{2}}\right) \\
c_{ij}^{(2)} &=&\frac{1}{k^{2}}\bigg\{ q_{i}q_{j}-f^{2}\delta
_{ij}-2(q_{i}k_{j}+q_{j}k_{i})\frac{{\bf q}\cdot {\bf k}}{k^{2}}
\\
&&+ \frac{k_{i}k_{j}}{k^{2}}\left[ f^{2}-{\bf q}^{2}
+\frac{4({\bf q}\cdot {\bf k})^{2}}{k^{2}}\right] \bigg\} \\
c_{ij}^{(3)} &=&\frac{1}{k^{4}}\bigg\{ -2{\bf q}\cdot {\bf k}
(q_{i}q_{j}-f^{2}\delta _{ij}) \\
&&+ (q_{i}k_{j}+q_{j}k_{i})\left[ f^{2}-{\bf q}^{2}
+\frac{4({\bf q}\cdot {\bf k})^{2}}{k^{2}}\right] \\
&& -4\frac{k_{i}k_{j}}{k^{2}} {\bf q}\cdot {\bf k}\left[
f^{2}-{\bf q}^{2} +\frac{2({\bf q}\cdot {\bf
k})^{2}}{k^{2}}\right] \bigg\} .
\end{eqnarray*}
This decomposition, besides separating the terms which give rise
to divergent contributions to $S_{ij}$ and $I_{ij}$, leads
naturally to an asymptotic expansion of these quantities for long
wavelengths. In fact, if we suppose that the refraction index
$n=c\kappa /\omega =q/f$ is of order unity (as it is reasonable to
expect for frequencies not too close to resonance), we have
$f\simeq q=a/\lambda $, and it is immediate to check that
$c_{ij}^{(l)}=O\left( (a/\lambda )^{l}\right) $,
$\bar{c}_{ij}=O\left((a/\lambda )^{4}\right) $ for $\lambda
\rightarrow \infty $.

\subsection{The lowest-order term}

Introducing ${\bf y}_{{\bf n}}={\bf r}_{{\bf n}}/a$, we can rewrite
Eq.\ (\ref{recipro}) in terms of dimensionless quantities as
\begin{equation}
\sum_{{\bf n}}\exp (2\pi i{\bf k\cdot y}_{{\bf n}})=
\sum_{{\bf m}}\delta ^{3}({\bf k}-{\bf H}_{{\bf m}})\,.
\label{recipro2}
\end{equation}
Let us then denote with $\hat{F}$ the Fourier transform operator,
which transforms the generic function $f({\bf k})$ into the function
\[
[\hat{F} f] ({\bf x})=\int d^{3}{\bf k}\,e^{i{\bf k} \cdot {\bf
x}}f({\bf k})\,.
\]
If the function $f$ is continuous at all points
${\bf H}_{{\bf m}}$, we
have from Eq.\ (\ref{recipro2})
\begin{eqnarray}
\sum_{{\bf m}}f({\bf H}_{{\bf m}}) &=&\int d^{3}{\bf k}\,
f({\bf k})\sum_{{\bf m}}\delta ^{3}({\bf k}-{\bf H}_{{\bf m}})
\nonumber \\
&=&\int d^{3}{\bf k}\,f({\bf k})\sum_{{\bf n}}\exp (2\pi i {\bf
k\cdot y}_{{\bf n}}) \nonumber \\
&=&\sum_{{\bf n}} [\hat{F} f] (2\pi {\bf y}_{{\bf n}})\,.
\label{fourier}
\end{eqnarray}
Applying this formula and Eq.\ (\ref{gauss}) we obtain
\begin{eqnarray}
\sum_{{\bf m}}\exp \left( -\eta {\bf H}_{{\bf m}}^{2}\right)
&=&\left( \frac{\pi }{\eta }\right) ^{3/2}\sum_{{\bf n}}\exp (-\pi
^{2}{\bf y}_{{\bf n}}^{2}/\eta ) \nonumber \\
&\rightarrow& \left( \frac{\pi }{\eta } \right) ^{3/2}+O(\eta
^{\infty })  \label{fourier0}
\end{eqnarray}
for $\eta \rightarrow 0^{+}$, where $O(\eta ^{\infty })$ indicates
a term such that $\lim_{\eta \rightarrow 0^{+}}O(\eta ^{\infty
})/\eta ^{N}=0$ for any arbitrarily large $N$. Let us then
decompose $c_{ij}^{(0)}$ as
\[
c_{ij}^{(0)}({\bf k)}=\frac{\delta _{ij}}{3}+\frac{P_{ij}^{(2)}
({\bf k)}}{k^{2}}\,,
\]
where
\[
P_{ij}^{(2)}({\bf k)}=k_{i}k_{j}-\delta _{ij}k^{2}/3
\]
is a quadratic symmetric tensor satisfying
$\sum _{i}P_{ii}^{(2)}({\bf k)}=0$. Since
\begin{equation}
b_{ij}^{(0)}({\bf k},\eta )\equiv \frac{P_{ij}^{(2)}({\bf k})}
{k^{2}}e^{-\eta k^{2}}  \label{b0def}
\end{equation}
is discontinuous for ${\bf k}={\bf 0}$, we cannot directly
apply Eq.\ (\ref {fourier}) to evaluate its contribution
to the series $S_{ij}$. However, since
\begin{eqnarray*}
\int d^{3}{\bf k}\,e^{i{\bf k}\cdot {\bf x}-\eta k^{2}}
P_{ij}^{(2)}({\bf k)} &=&-\frac{\pi ^{3/2}}{4\eta
^{7/2}}P_{ij}^{(2)}({\bf x)} \exp (-{\bf x}^{2}/4\eta ) \\
&\equiv& g_{ij}({\bf x},\eta )\,,
\end{eqnarray*}
we can use Eq.\ (\ref{fourier}) to obtain
\begin{eqnarray*}
-\frac{d}{d\eta }\sum_{{\bf m}\neq {\bf 0}}b_{ij}^{(0)} ({\bf
H}_{{\bf m}},\eta ) &=&\sum_{{\bf m}}P_{ij}^{(2)}({\bf H}_{{\bf
m}}) \exp \left( -\eta {\bf H}_{{\bf m}}^{2}\right) \\
&=&\sum_{{\bf n}}g_{ij}(2\pi {\bf y}_{{\bf n}},\eta )=O(\eta
^{\infty })\,.
\end{eqnarray*}
It follows that
\begin{equation}
\sum_{{\bf m}\neq {\bf 0}}b_{ij}^{(0)}({\bf H}_{{\bf m}},\eta )
\rightarrow \beta _{ij}^{(0)}+O(\eta ^{\infty })  \label{beta0}
\end{equation}
for $\eta \rightarrow 0^{+}$, where $\beta _{ij}^{(0)}$\ is a finite
symmetric dimensionless tensor, satisfying the
condition $\sum _{i}\beta_{ii}^{(0)}=0$. The numerical values of
$\beta_{ij}^{(0)}$ in general depend
on the particular crystal structure considered.
In Appendix A we derive the following equivalent expression in
terms of a sum over the points of the direct lattice:
\begin{equation}
\beta_{ij}^{(0)}=-\frac{3}{4\pi }\lim_{\tau \rightarrow 0^{+}}
\sum_{{\bf n}\neq {\bf 0}}
\frac{P_{ij}^{(2)}({\bf y}_{{\bf n}})}
{|{\bf y}_{{\bf n}}|^{5}}
\exp (-\tau {\bf y}_{{\bf n}}^{2})\,.  \label{beta0x}
\end{equation}
From Eqs.\ (\ref{fourier0}) and (\ref{beta0}) we conclude
\begin{eqnarray}
S_{ij}^{(0)}&\equiv& \sum_{{\bf m}\neq {\bf 0}}c_{ij}^{(0)} ({\bf
H}_{{\bf m}})\exp \left( -\eta {\bf H}_{{\bf m}}^{2}\right)
\nonumber \\
&\rightarrow& \frac{\delta_{ij}}{3} \left[ \left( \frac{\pi }{\eta
}\right) ^{3/2}-1\right] +\beta_{ij}^{(0)}+O(\eta ^{\infty })\,.
\label{c0s}
\end{eqnarray}
On the other hand, it is almost immediate to see that
\begin{equation}
\int d^{3}{\bf k}\,b_{ij}^{(0)}({\bf k},\eta )=0\,.  \label{b0}
\end{equation}
In fact, for obvious symmetry reasons, the l.h.s.\ must be of the
form $f(\eta)\delta _{ij}$,
but then the condition $\sum_{i}b_{ii}^{(0)}=0$ implies
$f(\eta )=0$. It follows that
\begin{eqnarray}
I_{ij}^{(0)}&\equiv& \int d^{3}{\bf k\,}c_{ij}^{(0)}({\bf k)} \exp
(-\eta k^{2}) \nonumber \\
&=&\frac{\delta _{ij}}{3}\int d^{3}{\bf k\,}\exp (-\eta k^{2})=
\frac{\delta _{ij}}{3}\left( \frac{\pi }{\eta }\right)^{3/2}
\label{c0i}
\end{eqnarray}
and so
\[
\lim_{\eta \rightarrow 0^{+}}\left( S_{ij}^{(0)}-I_{ij}^{(0)}\right)
=-\frac{\delta _{ij}}{3}+\beta _{ij}^{(0)}\,.
\]
We thus conclude that at order zero in $a/\lambda$ the tensor
$M^{{\rm ret}}_{ij}$ is given by
\begin{eqnarray}
M^{(0)}_{ij}({\bf q},f) &=& M^{\rm mac}_{ij}({\bf q},f)+
\frac{\delta _{ij}}{3}-\beta _{ij}^{(0)} \nonumber \\
&=& \frac{(2f^{2}+{\bf q}^{2})\delta_{ij}/3-q_{i}q_{j}} {{\bf
q}^{2}-f^{2}}-\beta _{ij}^{(0)}\,. \label{m0}
\end{eqnarray}

For an isotropic crystal the optical behavior at long wavelengths must
be invariant under spatial rotations. This means that
the symmetric tensor $\beta^{(0)}_{ij}$ must be a multiple
of the identity matrix, but since its trace
$\sum_{i}\beta_{ii}^{(0)}$ vanishes, we see that isotropy implies
$\beta _{ij}^{(0)}=0$.
In such a case, recalling Eqs.\ (\ref{pmacr}) and
(\ref{emacro2}), we derive from Eq.\ (\ref{m0}) that at this order of
approximation
\begin{equation}
{\bf e}_{{\bf n}}({\bf r}_{{\bf n}},t)={\bf E}_{{\rm mac}}
({\bf r}_{{\bf n}},t)+\frac 1 3 {\bf P}
({\bf r}_{{\bf n}},t)\,. \label{erregmacr}
\end{equation}
The above expression for the exciting field is the same that was
obtained by Lorentz and Planck with the argument of the virtual
sphere mentioned in the first section, and was later confirmed
through more rigorous analysis by Ewald and Born.

The same matrix ${\bf \hat{M}}^{(0)}$ of Eq.\ (\ref{m0}) of course
also provides the zeroth order approximation for the tensor ${\bf
\hat{M}}^{(+)}=V{\bf \hat{L}}^{(+)}$. Therefore recalling Eq.\
(\ref{eqmotoel}) one obtains
\begin{equation}
\frac{\omega^2_0-\omega^2}{\omega_p^2}{\bf P}=
{\bf E}_{\rm mac}+\frac 13 {\bf P}-\bm{\hat{\beta}}^{(0)}
\cdot {\bf P}\,, \label{eqmacr}
\end{equation}
where $\omega _{p}=e/\sqrt{mV}$ is the so-called ``plasma frequency''
of the material.
The above equation can be put into the form ${\bf P}=
\bm{\hat{\chi}}(\omega)\cdot {\bf E}_{\rm mac}$, where the tensor
\begin{equation}
\bm{\hat{\chi}}(\omega)=\left(\frac{\omega^2_0-\omega^2
-\omega_p^2/3}{\omega_p^2}{\bf \hat{1}}
+\bm{\hat{\beta}}^{(0)}\right)^{-1} \label{chi}
\end{equation}
represents the electric susceptibility. The dielectric function
can then by obtained as $\bm{\hat{\varepsilon}}(\omega)=
{\bf \hat{1}}+\bm{\hat{\chi}}(\omega)$. It appears from Eq.\ (\ref{chi})
that at the present order of approximation the lattice of resonators
behaves in general as a biaxial crystal.
Let us denote with $\bar{\beta}_i^{(0)}$ ($i=1,2,3$)
the three (in general distinct) eigenvalues of the
symmetric tensor $\bm{\hat{\beta}}^{(0)}$,
with
\[
\bar{\varepsilon}_i(\omega)
=1+\left(\frac{\omega^2_0-\omega^2
-\omega_p^2/3}{\omega_p^2}+\bar{\beta}_i^{(0)}\right)^{-1}
\]
the eigenvalues of $\bm{\hat{\varepsilon}}(\omega)$,
and with $n_i(\omega)=\sqrt{\bar{\varepsilon}_i(\omega)}$ the
indexes of refraction for transversal waves with electric fields
polarized along the mutually orthogonal directions of the
corresponding eigenvectors. We then have that, for $i\neq j$,
the (approximately) frequency independent quantities
\[
D_{ij}=\frac{1}{n_i^2-1}-\frac{1}{n_j^2-1}\,,
\]
that were introduced by Havelock \cite{havelock} as a measure of
the phenomenon of structural double refraction, can be calculated
according to our model as
\begin{equation}
D_{ij}=\bar{\beta}_i^{(0)}-\bar{\beta}_j^{(0)}\,.  \label{dij}
\end{equation}
According to Eqs.\ (\ref{beta0}) or (\ref{beta0x}) these parameters
depend solely on the lattice structure.

As we mentioned in Section I, Ewald was able to devise a method to
calculate $D_{ij}$ when the unit cell of the lattice is a
rectangular parallelepiped \cite{ewald1}. Although his formulas
look considerably more complicated, we have checked that the
numerical results, that he obtained for particular values of the
ratio between the edges of the cell, are in very good agreement
(considering the tools available at that time for numerical
computation) with the general result provided by Eq.\ (\ref{dij}).

In the following subsection we are going to show that Eq.\
(\ref{eqmacr}) has to be significantly corrected when the size of the
lattice parameter is not negligible with respect to the
wavelength. It will be found that it is no longer possible in that
case to give for the dielectric tensor a general expression as a
function of the frequency only, since, as was already recognized
by Ewald \cite{ewald1}, the form of the dispersion relation
involves both the direction of polarization and the direction of
the wavevector $\bm{\kappa}$. Our analysis will allow us to study
quantitatively in detail the behavior of such a dispersion
relation.

\subsection{The complete solution}

Since $c_{ij}^{(1)}$, as well as $c_{ij}^{(3)}$, are odd functions of
${\bf k}$, they clearly give no contributions to either $S_{ij}$ or
$I_{ij}$. Let us now consider $c_{ij}^{(2)}$.
If we introduce the fourth order completely symmetric tensor
\begin{widetext}
\begin{eqnarray*}
P_{ijhl}^{(4)}({\bf k})
&=&k_{i}k_{j}k_{h}k_{l}-\frac{k^{2}}{7}(\delta
_{ij}k_{h}k_{l}+\delta _{ih}k_{j}k_{l}+\delta
_{il}k_{j}k_{h}+\delta
_{jh}k_{i}k_{l}+\delta _{jl}k_{i}k_{h}+\delta _{hl}k_{i}k_{j}) \\
&&+\frac{k^{4}}{35}(\delta _{ij}\delta _{hl}+\delta _{ih}\delta
_{jl} +\delta_{il}\delta _{jh})
\end{eqnarray*}
satisfying the condition $\sum_{i}P_{iihl}^{(4)}({\bf k})=0$, we
obtain after some simple algebraic manipulation
\begin{eqnarray*}
c_{ij}^{(2)}({\bf k},{\bf q},f) &=&\frac{1}{15k^{2}}\left[
3q_{i}q_{j}-\delta _{ij}({\bf q}^{2}+10f^{2})\right] \\
&&+\frac{4}{7k^{4}}\left[ 4\delta _{ij}P_{hl}^{(2)}({\bf k})
q_{h}q_{l}-6q_{h}\left( q_{i}P_{jh}^{(2)}({\bf
k})+q_{j}P_{ih}^{(2)}
({\bf k})\right) +P_{ij}^{(2)}({\bf k})(7f^{2}-3{\bf q}^{2})\right] \\
&&+\frac{4}{k^{6}}P_{ijhl}^{(4)}({\bf k})q_{h}q_{l}\,.
\end{eqnarray*}
\end{widetext}
By integrating Eq.\ (\ref{fourier0}) with respect to $\eta $ we obtain
\begin{equation}
\sum_{{\bf m}\neq {\bf 0}}\frac{\exp \left( -\eta {\bf H}_{{\bf m}}^{2}
\right)}{{\bf H}_{{\bf m}}^{2}}=\frac{2\pi ^{3/2}}{\sqrt{\eta }}
-\alpha +\eta +O(\eta ^{\infty })\,,  \label{alpha}
\end{equation}
where $\alpha $ is a finite dimensionless constant dependent on the crystal
structure. Similarly, if we put
\[
b_{ij}^{(2)}({\bf k},\eta )\equiv \frac{P_{ij}^{(2)}({\bf k})}{k^{4}}%
e^{-\eta k^{2}},
\]
integration of Eq.\ (\ref{beta0}) provides
\begin{equation}
\sum_{{\bf m}\neq {\bf 0}}b_{ij}^{(2)}({\bf H}_{{\bf m}},\eta )\rightarrow
\beta _{ij}^{(2)}-\eta \beta _{ij}^{(0)}+O(\eta ^{\infty })\,,  \label{beta2}
\end{equation}
where $\beta _{ij}^{(2)}$\ is another finite symmetric tensor such that
$\sum_{i}\beta_{ii}^{(2)}=0$. Let us then define
\[
b_{ijhl}^{(n)}({\bf k},\eta )\equiv \frac{P_{ijhl}^{(4)}({\bf k})}{k^{4+n}}
e^{-\eta k^{2}}.
\]
We have
\begin{eqnarray*}
\int d^{3}{\bf k}\,e^{i{\bf k}\cdot {\bf x}-\eta
k^{2}}P_{ijhl}^{(4)} ({\bf k})&=&-\frac{\pi ^{3/2}}{16\eta
^{11/2}}P_{ijhl}^{(4)}({\bf x}) \exp (-{\bf x}^{2}/4\eta ) \\
&\equiv& g_{ijhl}({\bf x},\eta )
\end{eqnarray*}
whence, using Eq.\ (\ref{fourier}),
\begin{eqnarray*}
-\frac{d^{3}}{d\eta ^{3}}\sum_{{\bf m}\neq {\bf 0}}b_{ijhl}^{(2)}
({\bf H}_{{\bf m}},\eta )&=& \sum_{{\bf m}}P_{ijhl}^{(4)} ({\bf
H}_{{\bf m}})\exp \left( -\eta {\bf H}_{{\bf m}}^{2}\right) \\
&=&\sum_{{\bf n}}g_{ijhl}(2\pi {\bf y}_{{\bf n}},\eta )=O(\eta
^{\infty })\,.
\end{eqnarray*}
It follows that
\begin{equation}
\sum_{{\bf m}\neq {\bf 0}}b_{ijhl}^{(2)}({\bf H}_{{\bf m}},\eta)
=\gamma _{ijhl}^{(2)}-\gamma _{ijhl}^{(0)}\eta +\gamma
_{ijhl}^{(-2)}\eta ^{2}/2+O(\eta ^{\infty })\,,  \label{gamma2}
\end{equation}
where the integration constants $\gamma _{ijhl}^{(n)}$ are completely
symmetric tensors satisfying the condition $\sum_{i}\gamma_{iihl}^{(n)}
=0$ for $n=0,\pm 2$. We show in Appendix A that
\begin{eqnarray}
\gamma _{ijhl}^{(0)} &=&\frac{15}{8\pi }\lim_{\tau \rightarrow 0^{+}}
\sum_{{\bf n}\neq {\bf 0}}\frac{P_{ijhl}^{(4)}({\bf y}_{{\bf n}})}
{|{\bf y}_{{\bf n}}|^{7}}\exp (-\tau {\bf y}_{{\bf n}}^{2})  \label{g0ijhl} \\
\gamma _{ijhl}^{(-2)} &=&\frac{105}{16\pi ^{3}}\sum_{{\bf n}\neq {\bf 0}}
\frac{P_{ijhl}^{(4)}({\bf y}_{{\bf n}})}{|{\bf y}_{{\bf n}}|^{9}}\,.
\label{gm2ijhl}
\end{eqnarray}
A preliminary determination of $\gamma _{ijhl}^{(0)}$ and
$\gamma_{ijhl}^{(-2)}$ with the aid of the two above formulas can
considerably improve the accuracy in the numerical calculation of
$\gamma _{ijhl}^{(2)}$ according to Eq.\ (\ref{gamma2}). From
Eqs.\ (\ref{alpha}), (\ref{beta2}) and (\ref{gamma2}) it follows
that
\begin{eqnarray*}
S_{ij}^{(2)}&\equiv &\sum_{{\bf m}\neq {\bf 0}}c_{ij}^{(2)}({\bf H}_
{{\bf m}},{\bf q},f)\exp \left( -\eta {\bf H}_{{\bf m}}^{2}\right) \\
&\rightarrow &\frac{2}{15}\frac{\pi ^{3/2}}
{\sqrt{\eta }}\left[ 3q_{i}q_{j}-\delta _{ij}({\bf q}%
^{2}+10f^{2})\right] \\
&&-M_{ij}^{(2)}({\bf q},f)+O(\eta )\,,
\end{eqnarray*}
with
\begin{eqnarray}
M_{ij}^{(2)}({\bf q},f) &=&\frac{\alpha }{15}\left[ 3q_{i}q_{j}-
\delta _{ij}({\bf q}^{2}+10f^{2})\right]  \nonumber \\
&&-\frac{4}{7}\Big[ 4\delta _{ij}\beta
_{hl}^{(2)}q_{h}q_{l}-6q_{h} \left(q_{i} \beta_{jh}^{(2)}+q_{j}
\beta _{ih}^{(2)}\right) \nonumber \\
&&+\beta _{ij}^{(2)}(7f^{2}-3{\bf q}^{2})\Big] -4\gamma
_{ijhl}^{(2)}q_{h}q_{l}\,.  \label{m2}
\end{eqnarray}
By the same argument used to deduce Eq.\ (\ref{b0}) we have
\begin{equation}
\int d^{3}{\bf k}\,b_{ij}^{(2)}({\bf k},\eta )=0\,.
\end{equation}
In a similar way one finds that
\begin{equation}
\int d^{3}{\bf k}\,b_{ijhl}^{(2)}({\bf k},\eta )=0\,.  \label{bijhl}
\end{equation}
In fact, since the l.h.s.\ is an invariant tensor, it must be of the form
$f(\eta )(\delta _{ij}\delta _{hl}+\delta _{ih}\delta _{jl}+
\delta_{il}\delta _{jh})$, but then again the condition $\sum_{i}
b_{iihl}^{(2)}=0$ implies $f(\eta )=0$. It follows that
\begin{eqnarray*}
I_{ij}^{(2)} &\equiv &\int d^{3}{\bf k\,}c_{ij}^{(2)}({\bf k},{\bf q},f)
\exp (-\eta k^{2}) \\
&=&\left[ 3q_{i}q_{j}-\delta _{ij}({\bf q}^{2}+10f^{2})\right]
\int d^{3}{\bf k}\,\frac{\exp (-\eta k^{2})}{15k^{2}} \\
&=& \frac{2\pi^{3/2}}{15\sqrt{\eta }}\left[ 3q_{i}q_{j}-\delta
_{ij} ({\bf q}^{2}+10f^{2})\right] \,,
\end{eqnarray*}
so that
\[
\lim_{\eta \rightarrow 0^{+}}\left( S_{ij}^{(2)}-I_{ij}^{(2)}\right)
=-M_{ij}^{(2)}({\bf q},f)\,.
\]

Finally, we have
\begin{eqnarray*}
&&\lim_{\eta \rightarrow 0^{+}}\sum_{{\bf m}\neq {\bf 0}}
\bar{c}_{ij}({\bf H}_{{\bf m}},{\bf q},f) \exp \left( -\eta {\bf
H}_{{\bf m}}^{2}\right) \\
&=&\sum_{{\bf m}\neq {\bf 0}}\bar{c}_{ij}({\bf H}_{{\bf m}},{\bf
q},f) \equiv \bar{S}_{ij}({\bf q},f) \,, \\
&&\lim_{\eta \rightarrow 0^{+}}\int d^{3}{\bf k\,}\bar{c}_{ij}
({\bf k},{\bf q},f)\exp (-\eta k^{2}) \\
&=&\int d^{3}{\bf k\,}\bar{c}_{ij}({\bf k},{\bf q},f) \equiv
\bar{I}_{ij}({\bf q},f)\,,
\end{eqnarray*}
since both the sum and the integral on the r.h.s.\ of the two above
equations are absolutely convergent.
An explicit calculation carried out in Appendix B shows that
\begin{equation}
\bar{I}_{ij}({\bf q},f)=-i\frac{4}{3}\pi ^{2}f^{3}\delta _{ij}
=M^{(-)}_{ij}({\bf q},f)\,,  \label{ibar}
\end{equation}
in accordance with Eq.\ (\ref{oseen}).
We can therefore conclude that
\begin{equation}
M^{{\rm ret}}_{ij}({\bf q},f)=M^{(+)}_{ij}({\bf q},f)-i\frac{4}{3}\pi
^{2}f^{3}\delta _{ij}\,,  \label{mret}
\end{equation}
with
\begin{equation}
M^{(+)}_{ij}({\bf q},f)= M_{ij}^{(0)}({\bf q},f)
+M_{ij}^{(2)}({\bf q},f)-\bar{S}_{ij}({\bf q},f)\,.  \label{mpiu}
\end{equation}
As we had anticipated in the previous section, the matrix
$M^{(+)}_{ij}$ is symmetric and real for real ${\bf q}$ and $f$,
whereas $M^{(-)}_{ij}$ just cancels the radiation reaction term in
the equation of motion. Since $M^{(+)}_{ij}=VL^{(+)}_{ij}$, we can
rewrite Eq.\ (\ref{eqmotoel}) as
\[
{\bf \hat{M}}^{(+)}(\bm{\kappa}a/2\pi ,\omega a/2\pi c)\cdot {\bf C}=
\frac{\omega _{0}^{2}-\omega ^{2}}{\omega _{p}^{2}}{\bf C}
\]
or, in dimensionless variables,
\begin{equation}
\frac{r_{c}}{\pi a}{\bf \hat{M}}^{(+)}({\bf q},f)\cdot {\bf C}%
=(f_{0}^{2}-f^{2}){\bf C}\,,  \label{maineq}
\end{equation}
where $f_{0}=\omega _{0}a/2\pi c$, and
$r_{c}=e^{2}/4\pi mc^{2}$ is the classical electron radius.

\section{The simple cubic lattice}

In order to apply the general theory developed until now to a concrete
situation, let us consider in more detail the particular case of a
simple cubic lattice, for which the primitive translation vectors of
Eq.\ (\ref{lattice}) are given by ${\bf a}_{i}=a{\bf u}_{i}$ for
$i=1,2,3$, ${\bf u}_{i}$ being the unit vectors
of the three coordinate axes and $a$ the lattice
parameter. For this lattice one has ${\bf H}_{{\bf m}}={\bf m}$, and the
first Brillouin zone corresponds to the cubic region
$-1/2<q_{i}\leq 1/2$.
Obvious symmetry considerations imply that $\beta _{ij}^{(0)}=\beta
^{(0)}\delta _{ij}$, $\beta _{ij}^{(2)}=\beta ^{(2)}\delta _{ij}$. Then
from the general condition $\sum_{i}\beta_{ii}^{(0)}=\sum_{i}
\beta_{ii}^{(2)}=0$ one deduces that $\beta ^{(0)}=\beta ^{(2)}=0$,
and so
\[
\beta _{ij}^{(0)}=\beta _{ij}^{(2)}=0\,.
\]
The symmetry also implies that $\gamma _{ijhl}^{(n)}=\chi
^{(n)}E_{ijhl}-\gamma ^{(n)}D_{ijhl}$, where $D_{ijhl}\equiv\delta
_{ij} \delta_{hl}+\delta _{ih}\delta _{jl}+\delta _{il}\delta
_{jh}$ and the non-tensor $E_{ijhl}$\ is defined so that
$E_{ijhl}=1$ for $i=j=h=l$, $E_{ijhl}=0$ otherwise. Since
$\sum_{i}D_{iihl}=5\delta_{hl}$, $\sum_{i}E_{iihl}=\delta_{hl}$,
the condition $\sum_{i}\gamma_{iihl}^{(n)}=0$\ implies $\chi
^{(n)}=5\gamma ^{(n)}$. Equation (\ref{m2}) thus becomes
\begin{eqnarray*}
M_{ij}^{(2)}({\bf q},f)&=&\frac{\alpha }{15}\left[ 3q_{i}q_{j}-
\delta _{ij}({\bf q}^{2}+10f^{2})\right] \\
&& +4\gamma ^{(2)} (2q_{i}q_{j}+{\bf
q}^{2}\delta_{ij}-5q_{i}^{2}\delta _{ij})\,.
\end{eqnarray*}

A numerical calculation based on Eq.\ (\ref{alpha}) shows that
\[
\alpha =\lim_{\eta \rightarrow 0^{+}}\left( \frac{2\pi ^{3/2}}{\sqrt{\eta }}%
-\sum_{{\bf m}\neq {\bf 0}}\frac{e^{-\eta {\bf m}^{2}}}{{\bf m}^{2}}+\eta
\right) \cong 8.913633\,,
\]
while from Eqs.\ (\ref{gamma2}-\ref{gm2ijhl}) one obtains
\begin{widetext}
\begin{eqnarray*}
\gamma ^{(-2)} &=&\frac{1}{2}\gamma _{1111}^{(-2)}=\frac{21}{32\pi ^{3}}%
\sum_{{\bf n}\neq {\bf 0}}\frac{5n_{1}^{4}-|{\bf n}|^{4}}{|{\bf n}|^{9}}%
\cong 0.0751838 \\
\gamma ^{(0)} &=&\frac{3}{16\pi }\lim_{\tau \rightarrow 0^{+}}
\sum_{{\bf n}\neq {\bf 0}}\frac{5n_{1}^{4}-|{\bf n}|^{4}}{|{\bf n}|^{7}}
e^{-\tau {\bf n}^{2}}\cong 0.185800 \\
\gamma ^{(2)} &=&\frac{1}{10}\lim_{\eta \rightarrow 0^{+}}\left(
\sum_{{\bf m}\neq {\bf 0}}\frac{5m_{1}^{4}-|{\bf m}|^{4}}{|{\bf
m}|^{6}} e^{-\eta {\bf m}^{2}}+\gamma ^{(0)}\eta -\gamma
^{(-2)}\eta ^{2}/2\right) \cong 0.2780310\,.
\end{eqnarray*}
For application in numerical computations, the above results can also
be put in the useful form
\begin{eqnarray*}
M^{(+)}_{ij}({\bf q},f) &=&-\sum_{|{\bf m}|\leq L}
\frac{(m_{i}+q_{i})(m_{j}+q_{j})-f^{2}\delta _{ij}}
{({\bf m}+{\bf q})^{2}-f^{2}}+A(L)\frac{\delta _{ij}}{3} \\
&&+B(L)\left[ q_{i}q_{j}+({\bf q}^{2}-2f^{2})\delta _{ij}/3-2q_{i}^{2}
\delta_{ij}\right] -2C(L)(2q_{i}q_{j}+{\bf q}^{2}\delta _{ij}-5q_{i}^{2}
\delta_{ij}) \\
&&+\frac{4\pi }{105L}\left[ -q_{i}q_{j}(3{\bf q}^{2}-7f^{2})+
(|{\bf q}|^{4}+21{\bf q}^{2}f^{2}+70f^{4})\delta_{ij}\right]
+O(L^{-2})\,,
\end{eqnarray*}
\end{widetext}
where we have introduced the functions
\begin{eqnarray*}
A(L) &=&\sum_{|{\bf m}|\leq L}1 \\
B(L) &=&\sum_{{\bf m}\neq {\bf 0},\,|{\bf m}|\leq L}\frac{1}{{\bf m}^{2}}+\alpha \\
C(L) &=&\sum_{{\bf m}\neq {\bf 0},\, |{\bf m}|\leq
L}\frac{m_{1}^{4}}{|{\bf m}|^{6}}+\beta
\end{eqnarray*}
with $\beta =\alpha /5-2\gamma ^{(2)}\cong 1.226665$.

As an example of application of these formulas, we have numerically
calculated a few dispersion curves for an ideal simple
cubic lattice with parameter $a=2r_{B}$ and proper frequency of the
oscillators $\omega _{0}=\omega _{B}$,
where $r_{B}=4\pi \hbar^{2}/me^{2}$ and
$\omega _{B}=(e^{2}/4\pi )^{2}m/\hbar ^{3}$ are the Bohr radius
and the Bohr frequency respectively.
In Fig.\ 1 we show the dependence of $|{\bf q}|=\kappa a/2\pi$
as a function of $\omega/\omega_0$, as
determined according to Eq.\ (\ref{maineq})
for three different directions of propagation of the plane waves.
The results can be compared with
those obtained from the standard formula valid in the long wavelength limit,
which, as can be seen from Eqs.\ (\ref{maineq}) and (\ref{m0}), is given by
\[
\frac{r_{c}}{\pi a}\frac{(2f^{2}+{\bf q}^{2}){\bf C}/3-{\bf q}({\bf q}\cdot
{\bf C})}{{\bf q}^{2}-f^{2}}=(f_{0}^{2}-f^{2}){\bf C}\,.
\]
One can see that, according to the above equation,
${\bf C}$ has necessarily to be
either orthogonal or parallel to ${\bf q}$. In the first case
(transversal polarization) one obtains the dispersion relation
\[
\kappa =\frac{\omega }{c}\left( 1+\frac{\omega _{p}^{2}}{\omega
_{0}^{2}-\omega ^{2}-\omega _{p}^{2}/3}\right) ^{1/2},
\]
while the second case (longitudinal polarization) can occur only
when $\omega =\omega _{l}\equiv\sqrt{\omega _{0}^{2}+(2/3)\omega
_{p}^{2}}$, independently of the value of $\kappa$. These
relations are represented in the figure by the thick solid lines.
The curve corresponding to transversal polarization presents a
vertical asymptote for $\omega=\omega_{a}\equiv\sqrt{\omega
_{0}^{2}-\omega _{p}^{2}/3}$, whereas that for longitudinal
polarization is obviously just a vertical straight line at
$\omega=\omega_l$. According to the standard theory no propagation
is possible in the frequency interval between $\omega_a$ and
$\omega_l$. A further dispersion curve for transversal
polarization is present for $\omega>\omega_l$, according to both
the approximated and the exact theory, but its onset in the figure
is practically indistinguishable from the frequency axis. Note
that, for the particular numerical values that we have considered,
one has $r_{c}/\pi a=(\omega_{p}a/2\pi c)^{2}\cong 1/(2\pi \times
137^{2})$ and $f_{0}\cong 1/(\pi \times 137)$, so that $\omega
_{p}/\omega _{0}=\sqrt{\pi /2}\cong 1.253$,
$\omega_a/\omega_0=\sqrt{1-\pi/6}\cong 0.690$ and
$\omega_l/\omega_0=\sqrt{1+\pi/3}\cong 1.431$.

\begin{figure*}
\includegraphics{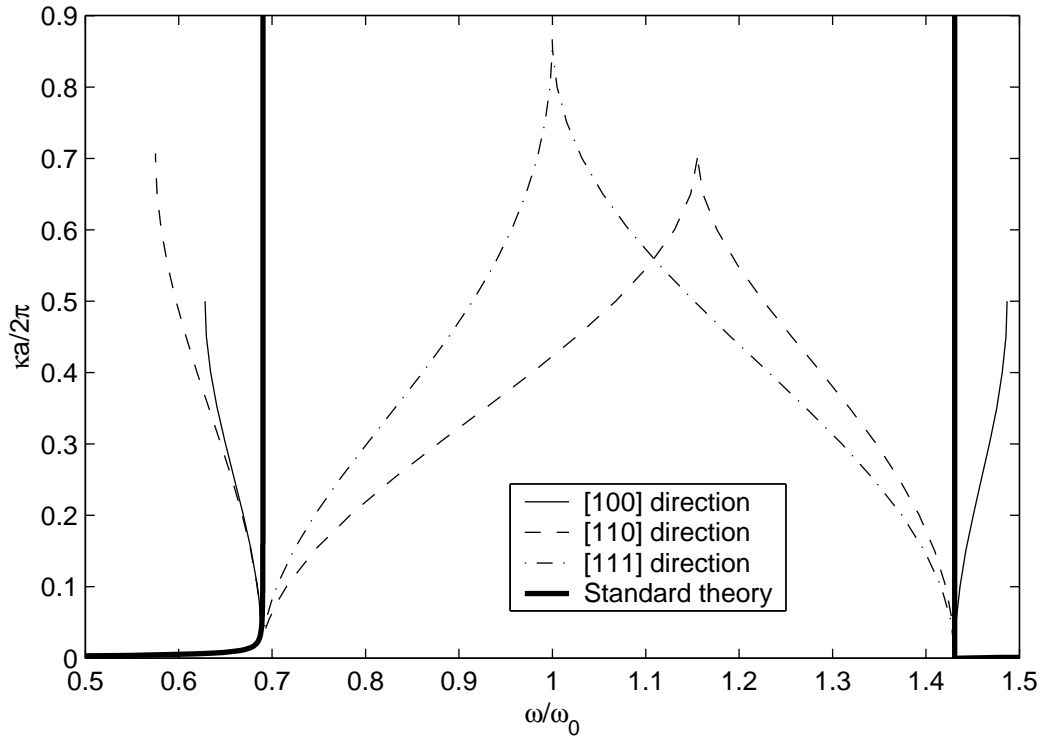}
\caption{Dispersion relation for three different directions of
propagation of a plane wave in a cubic lattice with $a=2r_B$ and
$\omega_0=\omega_B$. The polarization for each branch of these
curves is described in the text.}
\end{figure*}

As one could expect, one can notice that the predictions of our
exact theory depart in a significant way from those of the
approximated one as soon as $\kappa a/2\pi$ becomes of order of
magnitude comparable to unity. This situation of course
corresponds to the fact that the lattice parameter is no longer
negligible with respect to the wavelength. A new feature which is
revealed by the exact theory is the appearance, even for a perfect
cubic lattice, of an anisotropic behavior, consisting in a
remarkable dependence of the dispersion relation on the direction
of propagation and polarization of the plane waves. The two thin
solid curves at the opposite sides of the figure, which both refer
to the direction of propagation $[100]$, correspond to transversal
(the lower frequency one) and longitudinal polarization
respectively. If one considers the dependence of $\omega$ as a
function of $\kappa$ which is described by the former curve, one
can see that the frequency reaches a maximum --- which corresponds
to a zero of the group velocity $d\omega /d\kappa $ --- and then
slowly decreases over a wide region of the $\kappa$ axis extending
until the edge of the Brillouin zone. This means that there exists
a relatively small interval of the frequency axis with the
property that, for each $\omega $ belonging to this interval,
there are two transversal modes with direction of propagation
along a crystal axis and different values of the wavelength.

A similar phenomenon occurs also for the lowest-frequency branch
of the dashed curve, which corresponds to a wave again polarized
along a crystal axis, but propagating along the $[110]$ direction.
A completely different dispersion relation is shown instead by a
wave propagating in the same direction, but transversely polarized
along the $[1\bar{1}0]$ direction. This is illustrated by the
intermediate branch of the dashed line. For this curve the
frequency monotonically increases with $\kappa$ over the entire
Brillouin zone, but with a slope --- and therefore a group
velocity of the associated waves --- which drops to a relatively
small value when $\omega$ approaches and then exceeds $\omega_a$.
At the edge of the Brillouin zone this branch joins continuously,
at a point with vertical tangent, the third branch of the same
curve, which corresponds to longitudinal polarization. This branch
extends up to $\omega_l$, where a new transversal branch begins.
It follows that, at distinction with the modes propagating along
the $[100]$ direction, there is no forbidden frequency zone for
waves propagating along the $[110]$ direction. Of course, no hint
about the existence of such a phenomenon could have been derived
from the standard theory. A very similar behavior is shown by the
dash-dot curve, corresponding to waves propagating along the
$[111]$ direction. Here again the branch with transversal and that
with longitudinal polarization join continuously at the edge of
the Brillouin zone. Note that, whereas a wave propagating along
the $[110]$ direction can be transversely polarized along two
directions (the $[001]$ and $[1\bar{1}0]$) with distinct symmetry
properties and so also with different dispersion relations, the
two possible transversal polarizations are degenerate for waves
propagating along either the $[100]$ or $[111]$ directions.

For all the curves displayed in Fig.\ 1, the polarization vector
${\bf C}$ is either parallel
or orthogonal to the wavevector ${\bf q}$. We would like to point out
however that this fact is not a general property of the
the solutions of Eq.\ (\ref{maineq}), but is rather a consequence of
the special symmetry of the particular directions of propagation
that we have considered. Note also that the continuous links
between the transversal and the longitudinal branches of the curves
can be seen as an obvious consequence of the particular degeneracy
of the matrix ${\bf \hat{M}}^{(+)}({\bf q},f)$ when ${\bf q}=
(\frac 12,\frac 12,0)$ and ${\bf q}=(\frac 12,\frac 12,\frac 12)$.

\section{Concluding remarks}

In this paper we have presented a new refined version of the
classical dispersion theory of electromagnetic radiation in a
crystalline solid, which has been completely described at a
microscopic level as an infinite regular array of charged
oscillators, without ever making use of the continuum
approximation. It has been shown that, when the wavelength is
comparable with the lattice parameter, the predictions provided by
our calculations differ in a remarkable way from those derived
from the old standard approximated formulas. In particular, in the
region near the proper frequency of the oscillators, which was
believed to represent a forbidden gap for light propagation, the
phenomenon of ``slow light'' (i.e.\ light propagation with small
group velocity) is found instead to take place along appropriate
crystal directions.

The model we have studied is however interesting in itself also
from a purely theoretical point of view, independently of its
phenomenological implications. It is in fact an exceptional case
of a completely solvable model (although in the dipole
approximation) in which the radiation reaction force acting on
classical charged point particles is fully taken into account.
Actually, it turns out that the inclusion of this force has
fundamental implications on the qualitative properties of the
solutions, allowing for the existence of undamped collective
oscillations. Another interesting feature is that, although the
model was formulated in terms of the usual Maxwell--Lorentz
electrodynamics with Lorentz--Dirac selfinteraction, it is also
compatible with the action-at-a-distance electrodynamics which was
proposed by Wheeler and Feynman in 1945 \cite{WF}. In fact, when
the Wheeler--Feynman identity is verified, the two theories lead
to the same equations of motion for the charged particles.

Of course the investigations that we have presented here can be
developed and extended in several directions. One of these, which
appears to be particularly significant also from a
phenomenological point of view, is the study of the behavior of a
semi-infinite lattice occupying only one half of the full
three-dimensional space. This investigation should clarify the
relationship between the normal modes of the crystal --- which we
have described in the present work --- and the observable
electromagnetic radiation propagating in the free half-space,
i.e.\ the phenomena of refraction and reflection at the surface of
the crystal. In this way it is to be expected that classical
fundamental results, such as the Ewald--Oseen extinction theorem
\cite{oseen2,oseen3,ewald4,fearn}, will be justified on the basis
of the detailed microscopic dynamics of the complete system.

\appendix

\section{Series evaluation}

Let us introduce the operator $\hat{C}_{\tau }$, which acts on the generic
function $f({\bf k})$ by operating the convolution with a gaussian of
width $\sqrt{2\tau }$:
\[
[\hat{C}_{\tau }f] ({\bf p})=\int d^{3}{\bf k}\frac{\exp (-|{\bf
p}-{\bf k}|^{2}/4\tau )}{8(\pi \tau )^{3/2}}f({\bf k})\,.
\]
According to Eq.\ (\ref{gauss}), if $f$ is continuous in ${\bf
p}$, then $\lim_{\tau \rightarrow 0^{+}}[\hat{C}_{\tau } f] ({\bf
p})=f({\bf p})$. Furthermore we have for any $f$
\[
[\hat{F} \hat{C}_{\tau } f] ({\bf x})= e^{-\tau {\bf x}^{2}}
[\hat{F} f] ({\bf x})\,.
\]

The function $b_{ij}^{(0)}({\bf k},\eta )$, defined by Eq.\ (\ref{b0def}),
is continuous for ${\bf k}\neq {\bf 0}$\ and we have
\[
[\hat{C}_{\tau } b_{ij}^{(0)}] ({\bf 0},\eta )=\int d^{3} {\bf
k}\frac{\exp (-{\bf k}^{2}/4\tau )}{8(\pi \tau
)^{3/2}}b_{ij}^{(0)} ({\bf k},\eta )=0\,.
\]
Therefore, applying Eq.\ (\ref{fourier}) to the function
$\hat{C}_{\tau} b_{ij}^{(0)}$, we can write
\begin{eqnarray}
&&\sum_{{\bf m}\neq {\bf 0}}b_{ij}^{(0)}({\bf H}_{{\bf m}},\eta)
\nonumber \\
&=&\lim_{\tau \rightarrow 0^{+}}\sum_{{\bf m}} [\hat{C}_{\tau}
b_{ij}^{(0)}] ({\bf H}_{{\bf m}},\eta ) \nonumber \\
&=&\lim_{\tau \rightarrow 0^{+}}\sum_{{\bf n}} [\hat{F}
\hat{C}_{\tau } b_{ij}^{(0)}]
(2\pi {\bf y}_{{\bf n}},\eta )  \nonumber \\
&=&\lim_{\tau \rightarrow 0^{+}}\sum_{{\bf n}}e^{-\tau {\bf
y}_{{\bf n}}^{2}} [\hat{F} b_{ij}^{(0)}] (2\pi {\bf y}_{{\bf
n}},\eta )\,. \label{app_a1}
\end{eqnarray}
Using the formula
\[
\frac{1}{(2\pi )^{3}}\int d^{3}{\bf k\,}\frac{e^{i{\bf k}\cdot
{\bf x}}}{{\bf k}^{2}}=\frac{1}{4\pi x}\,,
\]
with standard arguments it is easy to show that
\[
\frac{1}{(2\pi )^{3}} [\hat{F} b_{ij}^{(0)}] ({\bf x},\eta
)\rightarrow -\frac{3}{4\pi x^{5}}P_{ij}^{(2)}({\bf x})+O(\eta
^{\infty })
\]
for $\eta \rightarrow 0^{+}$, ${\bf x}\neq {\bf 0}$, while
according to Eq.\ (\ref{b0})
\[
[\hat{F} b_{ij}^{(0)}] ({\bf 0},\eta )= \int d^{3}{\bf
k}\,b_{ij}^{(0)}({\bf k},\eta )=0\,.
\]
Therefore from Eq.\ (\ref{app_a1}) one deduces immediately Eqs.\ (\ref{beta0})
and (\ref{beta0x}). In a very similar way, by the same argument we used to
deduce Eq.\ (\ref{bijhl}) we have
\begin{eqnarray*}
[\hat{C}_{\tau } b_{ijhl}^{(0)}] ({\bf 0},\eta ) &=&0 \\
\small[\hat{F} b_{ijhl}^{(0)}\small] ({\bf 0},\eta ) &=&0\,,
\end{eqnarray*}
whereas one can show that
\begin{eqnarray*}
\frac{1}{(2\pi )^{3}} [\hat{F} b_{ijhl}^{(0)}] ({\bf x},\eta)
&\rightarrow& \frac{15}{8\pi x^{7}}P_{ijhl}^{(4)}({\bf x})\left(
1-14 \frac{\eta
}{x^{2}}\right) \\
&& +O(\eta ^{\infty })
\end{eqnarray*}
for $\eta \rightarrow 0^{+}$, ${\bf x}\neq {\bf 0}$. Applying
again Eq.\ (\ref{fourier}) we then have
\begin{eqnarray}
\sum_{{\bf m}\neq {\bf 0}}b_{ijhl}^{(0)}({\bf H}_{{\bf m}},\eta )
&=&\lim_{\tau \rightarrow 0^{+}}\sum_{{\bf m}}[\hat{C}_{\tau }
b_{ijhl}^{(0)}] ({\bf H}_{{\bf m}},\eta ) \nonumber
\\
&=&\lim_{\tau \rightarrow 0^{+}}\sum_{{\bf n}}[\hat{F}
\hat{C}_{\tau } b_{ijhl}^{(0)}]
(2\pi {\bf y}_{{\bf n}},\eta )  \nonumber \\
&=&\lim_{\tau \rightarrow 0^{+}}\sum_{{\bf n}}e^{-\tau {\bf
y}_{{\bf n}}^{2}} [\hat{F} b_{ijhl}^{(0)}] (2\pi
{\bf y}_{{\bf n}},\eta ) \nonumber \\
&=&\gamma _{ijhl}^{(0)}-\gamma _{ijhl}^{(-2)}\eta +O(\eta ^{\infty
})\,,
\end{eqnarray}
with $\gamma _{ijhl}^{(0)}$\ and $\gamma _{ijhl}^{(-2)}$\ given by Eqs.\ (\ref
{g0ijhl}) and (\ref{gm2ijhl}) respectively. Hence, by integrating with
respect to $\eta $, Eq.\ (\ref{gamma2}) is finally obtained.

\section{Calculation of $\bar{I}_{ij}$}

We have
\begin{eqnarray}
\bar{I}_{ij}({\bf q},f) &=&\int d^{3}{\bf k\,}\bar{c}_{ij}({\bf
k},{\bf q},f) \nonumber \\
&=&\lim_{L\rightarrow \infty }\int_{k<L}d^{3}{\bf k}\,\left(c_{ij}
-c_{ij}^{(0)}-c_{ij}^{(-2)}\right)  \nonumber \\
&=&\lim_{L\rightarrow \infty }\bigg\{ \int_{k<L}d^{3}{\bf
k}\,c_{ij}({\bf k}, {\bf q},f)-\frac{4\pi }{9}L^{3}\delta _{ij}
\nonumber \\
&&-\frac{4\pi }{5}L\left[ q_{i}q_{j}-\frac{\delta _{ij}}{3}\left(
{\bf q}^{2}+10f^{2}\right) \right] \bigg\} . \label{cbar}
\end{eqnarray}
Let us put $q=|{\bf q}|$. Since $c_{ij}$ is a second-rank tensor, from
invariance considerations it is readily found that we must have
\begin{eqnarray}
\int_{k<L}d^{3}{\bf k}\,c_{ij}({\bf k},{\bf q},f) &=& \frac{1}{2}
A(q,f)\left( \frac{3q_{i}q_{j}}{q^{2}} -\delta
_{ij}\right) \nonumber \\
&&-\frac{1}{2} B(q,f)\left( \frac{q_{i}q_{j}}{q^{2}} -\delta
_{ij}\right)  \label{celle}
\end{eqnarray}
with
\begin{eqnarray*}
A(q,f) &=&\int_{k<L}d^{3}{\bf k}\,c_{ij}({\bf k},{\bf q},f)
\frac{q_{i}q_{j}}{q^{2}} \\
B(q,f) &=&\int_{k<L}d^{3}{\bf k}\,c_{ij}({\bf k},{\bf q},f)
\delta_{ij}\,.
\end{eqnarray*}
After performing a rotation on the integration space ${\bf k}$, in such a
way that the unit vector ${\bf u}_{3}$\ of the third
coordinate axis is directed along the vector ${\bf q}$, we obtain
\begin{eqnarray}
A(q,f) &=&\int d^{3}{\bf k}\,\frac{k_{3}^{2}-f^{2}}{k^{2}-f^{2}-i\varepsilon
}\theta (L-|{\bf k}-q{\bf u}_{3}|)  \nonumber \\
&=&-f^{2}LI_{1}(q/L,f/L) \nonumber \\
&&+L^{3}I_{2}(q/L,f/L)  \label{aqf} \\
B(q,f) &=&\int d^{3}{\bf k}\,\frac{k^{2}-3f^{2}}{k^{2}-f^{2}
-i\varepsilon}
\theta (L-|{\bf k}-q{\bf u}_{3}|)  \nonumber \\
&=&\frac{4}{3}\pi L^{3}-2f^{2}LI_{1}(q/L,f/L)\,,  \label{bqf}
\end{eqnarray}
where
\begin{eqnarray*}
I_{1}(\bar{q},\bar{f}) &=&\int d^{3}{\bf k}\,\frac{\theta (1-|{\bf k}-
\bar{q}{\bf u}_{3}|)}{k^{2}-\bar{f}^{2}-i\varepsilon } \\
I_{2}(\bar{q},\bar{f}) &=&\int d^{3}{\bf k}\,\frac{k_{3}^{2}\theta (1-
|{\bf k}-\bar{q}{\bf u}_{3}|)}{k^{2}-\bar{f}^{2}-i\varepsilon }\,.
\end{eqnarray*}
We have
\begin{eqnarray}
B(0,f)&=&3A(0,f) \nonumber \\
&=&\frac{4}{3}\pi L^{3}-8\pi
f^{2}\left( L+f^{2}\int_{0}^{L}dk
\frac{1}{k^{2}-f^{2}-i\varepsilon }\right) \nonumber  \\
&\rightarrow& \frac{4}{3}\pi
L^{3}-8\pi f^{2}L-i4\pi ^{2}f^{3}  \label{b0f}
\end{eqnarray}
for $L\rightarrow \infty $, where we have used the relation
\begin{eqnarray*}
\int_{0}^{L}dk\frac{1}{k^{2}-f^{2}-i\varepsilon} &\rightarrow&
\frac{1}{2} \int_{-\infty }^{+\infty} dk\frac{1}{k^{2}-f^{2}-
i\varepsilon } \\
&&+O(L^{-1}) = \frac{\pi i}{2f}+O(L^{-1})\,.
\end{eqnarray*}
We have also for $\bar{q}=O(L^{-1})$, $\bar{f}=O(L^{-1})$%
\begin{eqnarray*}
\frac{\partial }{\partial \bar{q}}I_{1}(\bar{q},\bar{f}) &=&\int
d^{3} {\bf k}\,\frac{\delta (1-|{\bf k}-\bar{q}{\bf
u}_{3}|)}{k^{2}-\bar{f}^{2}} \frac{k_{3}-\bar{q}}{|{\bf
k}-\bar{q}{\bf u}_{3}|} \\
&=& \int d^{3}{\bf k} \,\frac{\delta
(1-k)}{k^{2}+2\bar{q}k_{3}+\bar{q}^{2}-\bar{f}^{2}}
\frac{k_{3}}{k} \\
&=&2\pi \int_{-1}^{1}d\xi \frac{\xi }{1+2\bar{q}\xi +\bar{q}^{2}-
\bar{f}^{2}} \\
&=& -\frac{8\pi }{3}\bar{q}+O(L^{-3})\,,
\end{eqnarray*}
whence
\begin{equation}
I_{1}(q/L,f/L)=I_{1}(0,f/L)+O(L^{-2})\,.  \label{i1}
\end{equation}
In a similar way we obtain
\begin{eqnarray*}
\frac{\partial }{\partial \bar{q}}I_{2}(\bar{q},\bar{f}) &=& 2\pi
\int_{-1}^{1}d\xi \frac{\xi (\xi +\bar{q})^{2}}{1+2\bar{q}\xi +
\bar{q}^{2}-\bar{f}^{2}} \\
&=& \frac{16\pi }{15}\bar{q}+O(L^{-3})\,,
\end{eqnarray*}
so that
\begin{equation}
I_{2}(q/L,f/L)=I_{2}(0,f/L)+\frac{8\pi }{15}\frac
{q^2}{L^2}+O(L^{-4})\,. \label{i2}
\end{equation}
From Eqs.\ (\ref{aqf}-\ref{i2}) we get
\begin{eqnarray*}
A(q,f) &=&\frac{4\pi }{9}L^{3}+\frac{8\pi }{3}\left(\frac {q^2}5-f^{2}\right)
L \\
&&-i\frac{4}{3}\pi ^{2}f^{3}+O(L^{-1}) \\
B(q,f) &=&\frac{4\pi }{3}L^{3}-8\pi f^{2}L-i4\pi ^{2}f^{3}+O(L^{-1})\,,
\end{eqnarray*}
then by substituting these expressions into Eqs.\ (\ref{celle})
and (\ref{cbar}) we finally obtain Eq.\ (\ref{ibar}).

\end{document}